\documentclass[12pt,a4paper]{article}

\usepackage{slashed}

\usepackage{a4wide}
\usepackage{latexsym}
\usepackage{epsf}
\usepackage{amssymb}
\usepackage{graphicx}
\usepackage{amsmath, cite}
\usepackage{amsmath,amssymb,amsthm, slashed}
\usepackage{booktabs,tabularx}
\usepackage{verbatim}
\usepackage{hyperref}
\usepackage{color}	
\usepackage{xcolor}
\usepackage{tcolorbox}
\usepackage{bbold}
\usepackage{comment}
\usepackage[normalem]{ulem}

\renewcommand{\d}{\textrm{d}}
\newcommand{\e}{\textrm{e}}

\newcommand{\be}{\begin{equation}}
\newcommand{\ee}{\end{equation}}
\newcommand{\beq}{\begin{equation}}
\newcommand{\eeq}{\end{equation}}
\newcommand{\ba}{\begin{eqnarray}}
\newcommand{\ea}{\end{eqnarray}}

\renewcommand{\d}{\textrm{d}}

\begin{document}
\numberwithin{equation}{section}

\begin{center}

{\LARGE Type IIA on $\mathrm{Spin}(7)$ manifolds with fluxes}

\vspace{2 cm} {\large   Niccol\`o Cribiori$^1$,  Arda Hasar$^2$, Thomas Van Riet$^1$, Timm Wrase$^3$}\\

\vspace{1.5 cm} 
{$^1$ Instituut voor Theoretische Fysica, KU Leuven,\\
Celestijnenlaan 200D B-3001 Leuven, Belgium}\\
\vspace{0.3 cm} 
{$^2$Department of Physics, Faculty of Arts and Sciences,\\ Middle East Technical University, 06800 Ankara, T\"urkiye}\\
\vspace{0.3 cm} {$^3$ Department of Physics, Lehigh University,\\
16 Memorial Drive East, Bethlehem, PA 18015, USA} \\

\vspace{3cm}
{\bf Abstract}
\end{center}

\begin{quotation}
We initiate a systematic study of compactifications of type II string theory to two dimensions on Ricci-flat spaces with fluxes and sources. We derive universal constraints on such compactifications, and then develop type IIA compactifications on $\mathrm{Spin}(7)$-holonomy spaces with bulk fluxes whose tadpole is cancelled by $\mathrm{OF1}$-planes. We derive the resulting two-dimensional $\mathcal N=(1,1)$ supergravity for a toroidal $\mathrm{Spin}(7)$ orbifold, and we extend the metric and universal sectors geometrically to general compact $\mathrm{Spin}(7)$ manifolds. In candidate supersymmetric Minkowski vacua, all untwisted shape modes appear in the flux scalar potential and can in principle be classically stabilised. However, in an explicit toroidal orbifold example flux quantisation together with the tadpole bound may obstruct the existence of candidate vacua supported entirely within the untwisted sector. The string-frame volume in string units is bounded by $7\chi/192$, so suppressing $\alpha'$ corrections requires $\mathrm{Spin}(7)$ manifolds with large Euler characteristic.
\end{quotation}

\newpage
\tableofcontents 
\newpage

\section{Introduction}
\label{sec:introduction}
Finding flux vacua of string theory is important for multiple reasons, including understanding the theory itself and its predictions. For phenomenological reasons, one would be tempted to only focus on four-dimensional (4D) vacua, but it is possible that also vacua with fewer than three large spatial dimensions may be relevant for understanding our four-dimensional world. The reason is that, in the past, the observable universe could have been in a phase where it was lower-dimensional \cite{Brandenberger:1988aj, Hertog:2021jyd} and this could imply a compact spatial topology of the current universe, perhaps at detectable scales \cite{Cornish:1997ab, COMPACT:2022gbl}. After all, to understand our world, one needs to do more than list all vacua and try to find our vacuum in this huge set: there can exist dynamical selection mechanisms that alter the number of visibly large dimensions \cite{Brandenberger:1988aj}. 

Another motivation for considering lower-dimensional flux vacua of string theory comes from holography. Indeed, finding AdS/CFT dual pairs is crucial for understanding both the space of CFTs and nonperturbative string theories. For this endeavour one is interested in flux vacua of any spacetime dimension, and particular attention has been paid to supersymmetric vacua, since supersymmetry strongly constrains perturbative instabilities and often improves control over possible nonperturbative decay channels.\footnote{It follows from Nahm's classification of superalgebras that supersymmetric AdS$_D$ vacua with the standard AdS superisometry algebra exist only for $D\leq 7$ \cite{Nahm:1977tg}.} Within the realm of supersymmetric vacua we have a particular interest in those with minimal supersymmetry. There are arguments suggesting that they are a promising setting for scale separation and full moduli stabilisation \cite{Cribiori:2022trc}, see however the recent paper \cite{Cribiori:2026caf}. 

In this work, we study minimal supersymmetric 2D flux vacua, which have received comparatively little attention in the literature; see nevertheless \cite{Cribiori:2024jwq, Macpherson:2024qfi} for recent work on anti-de Sitter and Minkowski flux compactifications. In 2D there are also general arguments suggesting that scale separation may be possible only with minimal supersymmetry \cite{Cribiori:2024jwq}, at most. Although scale separation was part of our initial motivation, in the class of compactifications considered here we do not find scale-separated AdS$_2$ vacua. Instead, we focus on classical Minkowski vacua.  Having a toroidal orbifold realisation in mind, we derive the associated flux scalar potential and its 2D $\mathcal N=(1,1)$ supergravity completion. The metric and universal sectors admit a geometric extension to general compact $\mathrm{Spin}(7)$ manifolds, while their axionic sectors require a separate derivation. The candidate supersymmetric Minkowski vacua fix in principle all propagating scalar modes, while the remaining constant volume-dilaton deformation is non-propagating and labels physically distinct backgrounds. For an explicit toroidal orbifold, however, and under certain working assumptions that we are going to specify, flux quantisation and the tadpole bound exclude candidate vacua supported entirely within the untwisted sector. Besides being conditional on our working assumptions, this result does not decide whether analogous vacua exist on other orbifolds or on generic smooth $\mathrm{Spin}(7)$ manifolds.

\section{Toroidal orbifolds, calibrations and orientifolds}\label{sec:2}

In this section, we discuss the link between G-structure forms (calibration forms) and brane intersections. This link was appreciated already a while ago \cite{Bergshoeff:1997kr,Gibbons:1998hm}, but it is particularly insightful for our context. The key lesson is that toroidal orbifolds allow for a simplified and intuitive treatment of several subtle aspects. 

\subsection{BPS brane/plane intersection rules}

We start with the standard rules for supersymmetric brane intersections in flat (10D) space. If two brane stacks are oriented in different orthogonal directions, the rule is that the number of directions with mixed Dirichlet and Neumann boundary conditions for strings has to be a multiple of 4. As an example, consider two stacks of D6-branes as in the table below.
\begin{center}
\begin{tabular}{ c || c c  c c c c c c c c}
& $x^0$ & $x^1$ & $x^2$ & $x^3$ & $x^4$ & $x^5$ & $x^6$ & $x^7$ & $x^8$ & $x^9$ \\  
\hline
D6$^A$ & $\star$ & $\star$ & $\star$ & $\star$ & $\star$ & $\star$ &  $\star$ & - & -  &-  \\
D6$^B$ & $\star$ & $\star$ & $\star$ & $\star$ & $\star$ & - & - & $\star$ & $\star$  & -   \\
\end{tabular}
\end{center}
A $\star$ means a direction along the brane stack and a - means perpendicular. The coordinates are labelled as $x^0\ldots x^9$; in the above diagram the directions $x^5, x^6, x^7$ and $x^8$ have a mixture of boundary conditions. One has to be careful when there are 8 such special directions, since then a (T-dual) version of the Hanany-Witten effect can take place \cite{Hanany:1996ie}. For example, a D0-D8 intersection cannot exist as such, but needs extra branes. This can be seen from the NSNS tadpole equation $d H_7 = F_0 F_8+\ldots$. A D0 brane would create $F_8$ flux and a D8 brane requires $F_0$. The term $F_0 F_8$ thus enters as a magnetic source for $H_7$, namely as the charge of a fundamental string (F1) stretching between the D0 and the D8, see the following table.
\begin{center}
\begin{tabular}{ c || c c  c c c c c c c c}
& $x^0$ & $x^1$ & $x^2$ & $x^3$ & $x^4$ & $x^5$ & $x^6$ & $x^7$ & $x^8$ & $x^9$ \\  
\hline
D0 & $\star$ & - & - & - & - & - &  -  & - & -  &-  \\
D8 & $\star$ & $\star$ & $\star$ & $\star$ & $\star$ & $\star$ & $\star$ & $\star$ & $\star$  & -   \\
F1 & $\star$ & - & - & - & - & - &  -  & - & -  & $\star$ \\
\end{tabular}
\end{center}

Below we will consider possible brane/plane intersections on toroidal orbifolds, needed for minimal supersymmetry in a given external spacetime. We will use that orientifolds obey the same BPS intersection rules as branes, and we will exploit the fact that toroidal orbifolds allow one to employ local Cartesian frames.
Aside from the phenomenological interest to reduce the amount of supersymmetry as much as possible, our motivation is ultimately the construction of scale-separated AdS vacua, for which strong arguments exist that they are unlikely with extended supersymmetry \cite{ Cribiori:2022trc, Montero:2022ghl, Cribiori:2023ihv, Cribiori:2024jwq}, except in 3D \cite{Cribiori:2026caf}.

\subsection{Compactifications to 4D}

Let us take type IIA string theory on an orbifold of $\mathbb{T}^6$ and try to put as many O6 planes as we can. The O6s have to be space-filling in the first four directions, $x^0,\ldots, x^3$, in order to have a maximally symmetric vacuum. The best we can do is depicted in the table below.\footnote{The $e^i=dx^{i+D-1}$, $i=1,2,\ldots, n$ are local vielbeins on the $n$-dimensional internal space with coordinates $x^{i+D-1}$; $D$ is the dimension of external space. We denote with $e^{ij}$ the wedge product $e^i\wedge e^j$.}
\begin{center}
\begin{tabular}{ c || c c  c  c | c c c c c c}
& $x^0$ & $x^1$ & $x^2$ & $x^3$ & $e^1$ & $e^2$ & $e^3$ & $e^4$ & $e^5$ & $e^6$ \\  
\hline
O6$^A$ & $\star$ & $\star$ & $\star$ & $\star$ & $\star$ & $\star$ &  $\star$ & - & -  & -  \\
O6$^B$ & $\star$ & $\star$ & $\star$ & $\star$ & $\star$ & - & - & $\star$ & $\star$  & -  \\
O6$^C$ & $\star$ & $\star$ & $\star$ & $\star$ & - & $\star$ & - & $\star$ & -  & $\star$  \\
O6$^D$ & $\star$ & $\star$ & $\star$ & $\star$ & - & - & $\star$ & - & $\star$  & $\star$  \\
\end{tabular}
\end{center}
This configuration is unique up to relabelling.  
In the internal space, any two O6s differ precisely along four directions, where one has a $\star$ and the other a -; in the remaining two directions either they both have $\star$ or they both have -. 
The product of two O6 orientifold involutions $\sigma$ gives an orbifold involution. For example, denoting by $\theta_{ijk\dots}$ the reflection of $e^i, e^j, e^k,...$, the product of $\sigma^A = \theta_{456}$ with $\sigma^B = \theta_{236}$ gives $\sigma^A \sigma^B=\theta_{2345}$. Conversely, multiplying an O6 orientifold involution by this orbifold involution gives another orientifold involution, whose fixed locus is the associated orientifold image, $i.e.$ another O6-plane in the diagram. For example, $\sigma^C \theta_{2345} = \theta_{135} \theta_{2345}=\theta_{124}=\sigma^D$, which follows from $\sigma^A \sigma^B \sigma^C \sigma^D = 1$ and $(\sigma^a)^2=1$ for $a= A,B,C,D$.

The diagram seems to suggest intersecting orientifolds, but it can be deceiving. On toroidal orbifolds, intersections of orientifold planes occur at orbifold singularities. When the orbifold is resolved to a smooth Calabi–Yau, these singularities are removed. Consequently, the apparent orientifold intersections are expected to disappear after resolution. This is illustrated in great detail in \cite{Junghans:2023yue}.

There is a nice link between the above intersection diagram and SU$(3)$-structures. On a $\mathbb{T}^6$, we have an SU$(3)$-invariant holomorphic $(3,0)$ form $\Omega=\Omega_R + i \Omega_I$, with real and imaginary parts such that $\Omega_R=\pm \star_6 \Omega_I$. The intersection diagram tells us that $\Omega_R$ is a linear combination of $e^{456}$, $e^{236}$, $e^{135}$ and $e^{124}$, but to fix the relative signs one needs extra information. Let us consider the combination $\Omega_R = a_1 e^{456} + a_2 e^{236} + a_3 e^{135} + a_4 e^{124}$, where without loss of generality $a_i = \pm 1$. Following \cite{Hitchin:2000jd}, it is convenient to introduce $\lambda(\Omega_R) = \frac{1}{6} K_{\Omega_R}^2$, where $K_{\Omega_R}$ is a certain quantity constructed out of $\iota_{e^i}\Omega_R \wedge \Omega_R$. As a consequence of  Proposition 2 of \cite{Hitchin:2000jd}, for $\Omega_R$ to give rise to an SU$(3)$-structure one needs that $\lambda(\Omega_R)<0$.
In our case, we can calculate that $\lambda(\Omega_R) \sim a_1 a_2 a_3 a_4$ and thus we need $a_1 a_2 a_3 a_4 <0$. One possible choice is
\begin{equation}
\Omega_R = e^{456} - e^{236} +e^{135} + e^{124}\,.    
\end{equation}

It is known that calibrated O6 planes inside an SU$(3)$-structure must have a source form proportional to $\Omega_R$, see $e.g.$ \cite{Koerber:2010bx}. In this sense, we can use the intersection diagram to predict the calibration forms for minimal supersymmetric configurations \cite{Bergshoeff:1997kr,Gibbons:1998hm}. Furthermore, as pointed out in \cite{Gukov:1999gr}, there is a natural pairing between the calibrated source forms and fluxes required to write down a superpotential $W$ for the lower-dimensional supergravity. Schematically, this reads
\begin{equation}
W \simeq \int \Phi \wedge F\,,    
\end{equation}
where $\Phi$ is the calibration form invariant under the structure group of the frame bundle, while $F$ is the flux (polyform) needed to cancel the tadpole of the sources. The integral then selects the top form out of the wedge product. In the example above, for O6 planes we need $H_3$ and $F_0$ flux to cancel the $dF_2$ tadpole, and the contribution to the 4D superpotential is proportional to the integral of $\Omega_I\wedge H_3 \sim \star \Omega_R \wedge H_3$. 

\subsection{Compactifications to 3D}

We now continue to compactifications of type IIA string theory on 7D manifolds preserving minimal supersymmetry: these are manifolds of $G_2$-holonomy. The framework for such flux compactifications was built in \cite{Farakos:2020phe, VanHemelryck:2022ynr}; see also the recent general treatment of $G_2$ flux compactifications in three dimensions in \cite{Aikot:2026tzn}. Following a similar reasoning, we arrive at the unique (up to relabelling) intersection diagram below. 
\begin{center}
\begin{tabular}{ c || c c  c | c  c c c c c c}
& $x^0$ & $x^1$ & $x^2$ & $e^1$ & $e^2$ & $e^3$ & $e^4$ & $e^5$ & $e^6$ & $e^7$ \\  
\hline
O6$^A$ & $\star$ & $\star$ & $\star$ & $\star$ & $\star$ & $\star$ &  $\star$ & - & -  & -  \\
O6$^B$ & $\star$ & $\star$ & $\star$ & $\star$ & $\star$ & - & - & $\star$ & $\star$  & -  \\
O6$^C$ & $\star$ & $\star$ & $\star$ & $\star$ & - & $\star$ & - & $\star$ & -  & $\star$  \\
O6$^D$ & $\star$ & $\star$ & $\star$ & $\star$ & - & - & $\star$ & - & $\star$  & $\star$  \\
O6$^E$ & $\star$ & $\star$ & $\star$ & - & - & $\star$ & $\star$ & $\star$ & $\star$  & -  \\
O6$^F$ & $\star$ & $\star$ & $\star$ & - & $\star$ & $\star$ & - & - & $\star$  & $\star$  \\
O6$^G$ & $\star$ & $\star$ & $\star$ & - & $\star$ & - & $\star$ & $\star$ & -  & $\star$  \\
\end{tabular}
\end{center}
This diagram tells us that the unique $G_2$-invariant 3-form is a linear combination of $e^{567}$, $e^{347}$, $e^{246}$, $e^{235}$, $e^{127}$, $e^{145}$ and $e^{136}$, but we need more information to find the right one. Let us thus consider the generic linear combination $\Phi_3 = a_1 e^{567} + a_2 e^{347} + a_3 e^{246} + a_4 e^{235}  + a_5 e^{127} +a_6e^{145} + a_7e^{136}$, with $a_i = \pm 1$. These signs can be restricted by demanding that the $G_2$ metric $g_{ij} \propto- \iota_{e^i}\Phi_3 \wedge \iota_{e^j} \Phi_3 \wedge \Phi_3$ be positive definite \cite{Hitchin:2000jd}. A direct calculation leads to the conditions $a_5 a_6 a_7 <0$, $a_3 a_4 a_5 >0$, $a_2 a_4 a_7<0$, $a_2 a_3 a_6>0$, $a_1 a_4 a_6<0$, $a_1 a_3 a_7>0$, $a_1 a_2 a_5<0$. One possible choice is $a_1=a_2=a_3=a_5=-1$, $a_4 =a_6=a_7=1$, leading to
\begin{equation}
\Phi_3 = - e^{567} - e^{347} - e^{246} + e^{235}  - e^{127} +e^{145} + e^{136},
\end{equation}
which up to relabelling and overall signs is the unique $G_2$-invariant 3-form. The 3D superpotential indeed contains the terms
\begin{equation}
\int_7 \Phi_3 \wedge F_4\,, \qquad \int_7 (\star_7 \Phi_3)\wedge H_3\,,   
\end{equation}
where $H_3$ is needed for the $O6$ tadpole (together with $F_0$) and $F_4$ is needed for $O2$ tadpoles (together with $H_3$), since the combined orientifold-orbifold group naturally induces $O2/O6$ bound states.  
Note that $O6^A, O6^B, O6^C$ and $O6^D$ are still the same as in the compactifications down to 4D. Going to 3D allowed us to add $O6^E, O6^F,$ and $O6^G$. 

\subsection{Compactifications to 2D}

Finally, we consider compactifications down to 2D. If we take our previous intersection diagram and simply promote one external direction to be compact, we find that there can be no more $O6$ planes added while respecting our rules. Indeed, we arrive at the intersection diagram reported below. 
\begin{center}
\begin{tabular}{ c || c c  | c  c  c c c c c c}
& $x^0$ & $x^1$  & $e^1$ & $e^2$ & $e^3$ & $e^4$ & $e^5$ & $e^6$ & $e^7$  & $e^8$\\  
\hline
O6$^A$ & $\star$ & $\star$ & $\star$ & $\star$ & $\star$ & $\star$ &  $\star$ & - & -  & -  \\
O6$^B$ & $\star$ & $\star$ & $\star$ & $\star$ & $\star$ & - & - & $\star$ & $\star$  & -  \\
O6$^C$ & $\star$ & $\star$ & $\star$ & $\star$ & - & $\star$ & - & $\star$ & -  & $\star$  \\
O6$^D$ & $\star$ & $\star$ & $\star$ & $\star$ & - & - & $\star$ & - & $\star$  & $\star$  \\
O6$^E$ & $\star$ & $\star$ & $\star$ & - & - & $\star$ & $\star$ & $\star$ & $\star$  & -  \\
O6$^F$ & $\star$ & $\star$ & $\star$ & - & $\star$ & $\star$ & - & - & $\star$  & $\star$  \\
O6$^G$ & $\star$ & $\star$ & $\star$ & - & $\star$ & - & $\star$ & $\star$ & -  & $\star$  \\
\end{tabular}
\end{center}
Notice also that the $e^1$-direction is singled out as the only direction inside the compact manifold that is wrapped by all $O6$ planes.  
This could mean that the associated volume modulus does not get stabilised when one tries to compactify on a Ricci-flat space of restricted holonomy, in the absence of fluxes. It could also be related to the lack of a natural top form on the same space. In any case, our empirical rule for BPS brane/plane intersections breaks down for 2D toroidal compactifications with orientifolds. The reason is that $O6$ planes are not the appropriate objects to use in this setup, and other planes are required, as we now explain. 

The special holonomy manifold of dimension $8$ needed for minimal supersymmetry is a $\mathrm{Spin}(7)$ manifold \cite{Berger:1955yyj}. The calibrated form is known as the Cayley form and is a 4-form. This means that the orientifold sources should be of codimension 4 and the flux needed should be a 4-form flux. Hence, to apply our rule for BPS brane/plane intersections in compactifications down to 2D we need 5-planes rather than 6-planes. On the one hand, this explains why the previous attempt was not successful, on the other hand we now face the problem that type IIA string theory has 4-form flux but no standard $O5$ planes, at least at first sight.\footnote{Once one allows for torsion in the compact manifold, $d e^i \neq 0$, one can create higher forms such as $d\Phi$ and pair them with fluxes. We do not consider this possibility here.} Luckily, this last statement is not quite correct. Besides the standard $O5$ planes, other 5 planes exist in type II \cite{Witten:1998xy,Hanany:2000fq}, some of which can be detected using string dualities \cite{Hanany:2000fq}. Certain of these $5$ planes are called $\mathrm{ON5}$ and they form natural bound states with so-called $\mathrm{OF1}$ planes that would be space-filling in 2D.  In the next subsection we explain in a bit more detail what these orientifold planes are, but we first consider the consequences of their existence in the context of calibrations and intersections in toroidal orbifolds.

We thus build the intersection diagram of $\mathrm{ON5}$ planes wrapping cycles in a compact 8D manifold and following our usual rules for BPS brane/plane intersections. This is the unique (up to relabelling) diagram reported below. 

\begin{center}
\begin{tabular}{ c || c c | c c c c c c c c}
& $x^0$ & $x^1$  & $e^1$ & $e^2$ & $e^3$ & $e^4$ & $e^5$ & $e^6$ & $e^7$  & $e^8$\\  
\hline
$\mathrm{ON5}^A$ & $\star$ & $\star$ & - & - & - & - & $\star$ & $\star$ & $\star$ & $\star$  \\
$\mathrm{ON5}^B$ & $\star$ & $\star$ & - & - & $\star$ & $\star$ & - & - & $\star$ & $\star$  \\
$\mathrm{ON5}^C$ & $\star$ & $\star$ & -  & - & $\star$ & $\star$ & $\star$ & $\star$ & - & - \\
$\mathrm{ON5}^D$ & $\star$ & $\star$ & -  & $\star$ & - & $\star$ & - & $\star$ & - & $\star$ \\
$\mathrm{ON5}^E$ & $\star$ & $\star$ & -  & $\star$ & - & $\star$ & $\star$ & - & $\star$ & - \\
$\mathrm{ON5}^F$ & $\star$ & $\star$ & -  & $\star$ & $\star$ & - & - & $\star$ & $\star$ & - \\
$\mathrm{ON5}^G$ & $\star$ & $\star$ & -  & $\star$ & $\star$ & - & $\star$ & - & - & $\star$ \\
$\mathrm{ON5}^{A'}$ & $\star$ & $\star$ & $\star$ & $\star$ & $\star$ & $\star$ & - & - & - & -  \\
$\mathrm{ON5}^{B'}$ & $\star$ & $\star$ & $\star$ & $\star$ & - & - & $\star$ & $\star$ & - & -  \\
$\mathrm{ON5}^{C'}$ & $\star$ & $\star$ & $\star$  & $\star$ & - & - & - & - & $\star$ & $\star$ \\
$\mathrm{ON5}^{D'}$ & $\star$ & $\star$ & $\star$  & - & $\star$ & - & $\star$ & - & $\star$ & - \\
$\mathrm{ON5}^{E'}$ & $\star$ & $\star$ & $\star$  & - & $\star$ & - & - & $\star$ & - & $\star$ \\
$\mathrm{ON5}^{F'}$ & $\star$ & $\star$ & $\star$  & - & - & $\star$ & $\star$ & - & - & $\star$ \\
$\mathrm{ON5}^{G'}$ & $\star$ & $\star$ & $\star$  & - & - & $\star$ & - & $\star$ & $\star$ & - \\
\end{tabular}
\end{center}

The second set of seven $\mathrm{ON5}$ planes is denoted with a prime to indicate that they wrap the cycles dual to those wrapped by the first seven. From this diagram, we see that the Cayley form should be given by a linear combination of the type $\Phi_4 = a_1 e^{1278} + a_2 e^{1368}+ a_3 e^{1458} + a_4 e^{5678} + a_5 e^{3478}+ a_6 e^{2468} + a_7 e^{2358}+a_8 e^{1234} + a_9 e^{1256} + a_{10}e^{1357} + a_{11}e^{1467} + a_{12}e^{3456} + a_{13}e^{2457} + a_{14}e^{2367}$, with $a_i = \pm 1$. If the 8D manifold is the product of a 7D manifold times an $S^1$, Example 5.3.7 of \cite{Karigiannis2005} tells us that $\Phi_4 = \varphi_3\wedge e^8 + \star_7 \varphi_3$ and $\Phi_4$ defines a $\mathrm{Spin}(7)$-structure if $\varphi_3$ defines a $G_2$-structure.  With our choice of basis, one can see that $\varphi_3$ coincides with $\Phi_3$ of the previous section and thus we can readily read from there the restrictions on the coefficients $a_i$ in order to have a $G_2$-structure.  Besides, one needs also to impose $\Phi_4 = \star_8 \Phi_4$, which relates $a_{3+k} = a_{7+k}$, $k=1,2,3,4$, and $a_1=a_{12}$, $a_{2} = a_{13}$, $a_3 = a_{14}$. Putting the two conditions together, we find the restrictions $a_1 a_2 a_3<0$, $a_1 a_6 a_7>0$, $a_2 a_5 a_7 <0$, $a_3 a_5 a_6 >0$, $a_3 a_4 a_7<0$, $a_2 a_4 a_6>0$, $a_1 a_4 a_5<0$. One possible choice is $a_4=a_5=a_6 =a_1 =-1$ and $a_7=a_2 =a_3=1$, leading to
\begin{equation}
\begin{aligned}
\Phi_4 = &- e^{1234} - e^{1256} - e^{1278} - e^{1357} + e^{1368} + e^{1458} + e^{1467} \\&+ \star_8(- e^{1234} - e^{1256} - e^{1278} - e^{1357} + e^{1368} + e^{1458} + e^{1467})\, ,
\end{aligned}
\label{eq:Phi4ei}
\end{equation}
consistently, up to an overall sign, with the known expression for the unique $\mathrm{Spin}(7)$-invariant 4-form in $\mathbb{R}^8$ \cite{Gauntlett:2003cy}.

Note that now the $\mathrm{ON5}$ planes together with their $\mathrm{ON5}'$ partners have $8$ mixed directions. Accordingly, we expect a Hanany-Witten brane creation effect. This can indeed be seen from the involutions themselves. Let us for instance consider $\mathrm{ON5}^A$ and $\mathrm{ON5}^{A'}$. The spacetime involutions are
\begin{align}
\theta_A: \quad e^{1,2,3,4}\rightarrow -e^{1,2,3,4}\, ,\\
\theta_{A'}: \quad e^{5,6,7,8}\rightarrow -e^{5,6,7,8}\, ,
\end{align}
and their product leads to
\begin{equation}
e^{1,2,3,4,5,6,7,8}\rightarrow -e^{1,2,3,4,5,6,7,8}\, ,
\end{equation}
which corresponds to an $\mathrm{OF1}$ action, as predicted. One can check that the whole table is consistent with any combination of such multiplications; for instance, $\theta_A \theta_B =\theta_{C'}$ and similarly for other entries. 
For the more conventional RR-charged orientifolds, we normally have that the product of two orientifold actions is an orbifold action and not an orientifold action since the worldsheet involution squares to one. Indeed, the $\mathrm{OF1}$ has no associated worldsheet parity action \cite{Hanany:2000fq} and is really an orbifold, as reviewed in the next section.

Later in this paper we derive the 2D effective theory of an orientifolded toroidal $\mathrm{Spin}(7)$ orbifold within this setup and verify that  the 2D superpotential does contain a term proportional to
\begin{equation}
 W \simeq \int_8 \Phi_4 \wedge F_4 \,,  
 \label{eq:Wcal2D}
\end{equation}
with $\Phi_4$ the Cayley 4-form. Although we derive it on a toroidal orbifold, the resulting 2D supergravity takes a manifestly geometric form valid on general $\mathrm{Spin}(7)$ manifolds, see section~\ref{sec:generic}; only the axionic sectors present when the Betti numbers $b_2,b_3\neq0$ require a separate derivation, which is left for future work. 

Note that $\mathrm{Spin}(7)$ manifold reductions have appeared in several places in the literature before. For example M-theory on a compact $\mathrm{Spin}(7)$ manifold yields 3D $\mathcal N=1$ vacua \cite{Becker:2000jc,Acharya:2002vs,Becker:2003wb,Majumder:2001dx}. The set-up in our current work, to our knowledge, is new.

\subsection{On exotic orientifolds}
\label{sec:commentsorientifolds}

The most common orientifold planes in string perturbation theory are fixed loci of worldsheet parity, $\Omega_p$, combined with a spacetime involution, $\theta$, and, in certain cases, with a $(-1)^{F_L}$ action. From the DBI and CS couplings of their effective actions, one can see that they carry negative tension together with an RR charge. 
In supersymmetric compactifications these planes can be calibrated sources and must be accompanied by the corresponding Bianchi identities, tadpole constraints and flux quantization conditions.

A finer classification involves additional discrete NS/RR charges valued in cohomology with (twisted) integer coefficients, as explained in \cite{Witten:1998xy, Hanany:2000fq}. 
The discrete charges allow for new orientifold variants, some of which are not visible in string perturbation theory. 
A more precise classification is given by K-theory \cite{Bergman:2001rp}, where cohomology charges re-organise and some of the variants may disappear. We will not need those topological considerations here, but we will make use of certain orientifolds which are somehow less standard from the viewpoint of perturbative string model building. Examples of these are the defects denoted $\mathrm{ON5}$ and $\mathrm{OF1}$ in the previous section. They can be found from the more standard orientifolds via string dualities.

A quick argument proceeds as follows \cite{Hanany:2000fq}. Consider the usual $O1$ plane in type IIB carrying electric RR $C_2$-charge; we denote it as $IIB/\theta_8 \Omega_p$, where $\theta_8$ is the inversion of eight spacetime coordinates. T-duality to type IIA maps it to either $O0$ or $O2$ planes. Instead, let us first consider its S-dual counterpart. Since S-duality exchanges $C_2$ with $B_2$, this object should carry electric $B_2$-charge, such as a fundamental string; on the other hand, its tension is negative. In the terminology of \cite{Hanany:2000fq}, this is an $\mathrm{OF1}_B$ plane.\footnote{This $\mathrm{OF1}_B$ comes with a certain number of variants depending on the values of certain discrete charges, or equivalently on certain brane intersections \cite{Witten:1998xy, Hanany:2000fq,Bergman:2001rp}. For our purposes the crucial point is not their full microscopic classification, but the fact that they are non-dynamical negative-tension defects with NS rather than RR charge and dilaton dependence in their tension. 
} Since S-duality exchanges $\Omega_p$ with $(-1)^{F_L}$ in type IIB, the $\mathrm{OF1}_B$ is really the orbifold $IIB/\theta_8 (-1)^{F_L}$, with no worldsheet parity action. 
Now, T-dualise to type IIA along a direction transverse to the fixed line. This operation is accompanied by a $(-1)^{F_L}$ action and thus, since $(-1)^{F_L}$ squares to 1, we reach the orbifold $IIA/\theta_8$. In the terminology of \cite{Hanany:2000fq} this is denoted as $\mathrm{OF1}_A$, and it is the $\mathrm{OF1}$ plane we mentioned in the previous section.
\begin{equation*}
O1 \equiv IIB/\theta_8 \Omega_p \quad \overset{S}{\longleftrightarrow} \quad \mathrm{OF1}_B \equiv IIB/\theta_8(-1)^{F_L}\quad \overset{T_{\perp}}{\longleftrightarrow} \quad \mathrm{OF1}_A \equiv IIA/\theta_8
\end{equation*}

The $\mathrm{OF1}_A$ used in section \ref{sec:toroidal_bulk_moduli} is therefore a pure spacetime orbifold rather than a conventional worldsheet-parity orientifold. In our toroidal orbifold model, the inversion $\theta_8$ will act on the internal spinor as the 8D chirality operator and will leave the unique positive-chirality $\mathrm{Spin}(7)$ singlet invariant. It will therefore impose no additional independent supersymmetry projection: the two real supercharges selected by the compactification will survive, giving $\mathcal N=(1,1)$ supergravity in 2D.

As for the $\mathrm{ON5}$, one can argue similarly \cite{Hanany:2000fq}. The starting point is now the usual $O5$ plane in type IIB carrying electric RR $C_6$-charge; we denote it as $IIB/\theta_4 \Omega_p$. S-duality brings us to the orbifold $IIB/\theta_4 (-1)^{F_L}$ which in the terminology of \cite{Hanany:2000fq} is called $\mathrm{ON5}_B$. It carries NS $B_6$-charge and has negative tension. T-duality along a direction parallel to the fixed plane gives the orbifold $IIA/\theta_4 (-1)^{F_L}$ which is termed $\mathrm{ON5}_A$.\footnote{Recall that $T$-duality along an $F1_{A/B}$ gives an $F1_{B/A}$. $T$-duality along an $NS5_{A/B}$ gives an $NS5_{B/A}$ while $T$-duality transverse to an $NS5_{A/B}$ gives a $KKm_{B/A}$, see $e.g.$\cite{Papadopoulos:1997je,Eyras:1998hn}.} This is the $\mathrm{ON5}$ plane we mentioned in the previous section.   
\begin{equation*}
O5 \equiv IIB/\theta_4 \Omega_p \quad \overset{S}{\longleftrightarrow} \quad \mathrm{ON5}_B \equiv IIB/\theta_4(-1)^{F_L}\quad \overset{T_{\parallel}}{\longleftrightarrow} \quad \mathrm{ON5}_A \equiv IIA/\theta_4(-1)^{F_L}
\end{equation*}
For our $\mathrm{Spin}(7)$ compactifications, the $(-1)^{F_L}$ acts with opposite sign on the two supercharges (the $\mathcal N=(1,1)$ above). Spacetime-filling $\mathrm{ON5}_A$ planes that wrap Cayley-calibrated 4-cycles will therefore project out one of the supercharges. This leaves a minimal supergravity with a single supercharge.

This reasoning tells us that the tension and NS charge of these less common orientifold planes can be read off from their supergravity effective actions. In particular, the tension of the $\mathrm{OF1}/\mathrm{ON5}$ has the same dilaton dependence as that of an $F1/NS5$, but opposite sign.

\section{2D flux vacua: universal fields}
\label{sec:universal_constraints}

Flux vacua in string theory are constrained already at tree level by simple scaling arguments involving fluxes, curvature and localised sources \cite{VanRiet:2023pnx}.
In compactifications down to $D\geq 3$ one usually formulates such arguments in the lower-dimensional Einstein frame. In 2D this route is not available, because the Einstein--Hilbert term is topological and does not produce dynamics. For this reason, it is more natural to keep the external length scale explicit, as we do in the following. 

\subsection{Reduced effective action}
\label{subsec:reduced_action_2d}

We consider the bosonic part of the ten-dimensional type II string theory action in Einstein frame. We use units \(2\pi\sqrt{\alpha'}=1\), so that
\(2\kappa_{10}^2=(2\pi)^7(\alpha')^4=(2\pi)^{-1}\), and hence the Einstein-frame action carries the overall prefactor \(2\pi\),
\begin{equation}
S_{10}=2\pi \int \d^{10}x\,\sqrt{-g_{10}}\left(R_{10}-\frac{1}{2} (\partial \phi)^2-\frac{1}{2}\e^{-\phi}H_3^2-\frac{1}{2}\sum_n \e^{a_n\phi}F_n^2\right)\, ,
\label{eq:10d_action_universal}
\end{equation}
with $a_n=\frac{5-n}{2}$. We use conventions where $H_3^2 = \frac{1}{3!} H_{MNP}H^{MNP}$ and similarly for $F_n^2$.
We take all fluxes purely internal and we temporarily switch off  localised sources.
The integer $n$ runs over the RR field strengths appropriate to the chosen type II theory. Alternatively, one can consider type I and heterotic compactifications by appropriately setting to zero certain fluxes.  

Compactification of the above action contains the string coupling, the volume of the internal manifold, and the external curvature scale, $L$, parametrising the metric. We will refer to these as \emph{universal fields}, and focus on maximally symmetric vacua in which they assume constant values. Their algebraic dependence in the reduced action is enough to constrain the allowed flux compactifications. 

To compactify the theory we take a direct-product Ansatz
\begin{equation}
\d s_{10}^2=L^2\,\d \tilde s_2^2+\rho\,\d \tilde s_8^2,
\label{eq:metric_ansatz_universal}
\end{equation}
where $\d \tilde s_2^2$ is a unit-radius maximally symmetric 2D metric, $\d \tilde s_8^2$ is a reference internal metric normalised to unit volume, $\int_8 \sqrt{\tilde g_8}=1$, and $\rho$ is the universal field associated to the volume of the internal manifold. We parameterise the external curvature by
\begin{equation}
R^{(2)}_{\mu\nu}=\frac{\lambda}{L^2}\, g^{(2)}_{\mu\nu},
\qquad \lambda\in\{-1,0,+1\},
\label{eq:lambda_definition_universal}
\end{equation}
so that the physical 2D scalar curvature is $R_2=2\lambda/L^2$. With these conventions, the 10D Ricci scalar splits as
\begin{equation}
R_{10}=\frac{2\lambda}{L^2}+\frac{\widetilde R_8}{\rho},
\label{eq:R10_split_universal}
\end{equation}
where $\widetilde R_8$ is the scalar curvature of the unit-volume reference metric.

After integrating over the internal manifold, the reduced action is given by
\begin{equation}
S_2 = 2\pi \int \d^2x\, \mathcal{L} \equiv 2\pi \int \d^2x\,\sqrt{-\tilde g_2}\,\Big(2\lambda\rho^4 +{\rm kin}- V_R-V_H-\sum_n V_{F_n} \Big),
\label{eq:S2_schematic_universal}
\end{equation}
where kin stands for the kinetic terms of scalar fields and where the curvature and flux terms scale as
\begin{align}
V_R &= -L^2\rho^3\,\widetilde R_8 ,
\label{eq:VR_universal}\\
V_H &= \frac{1}{2}L^2\rho\,\e^{-\phi}\,\widetilde H_3^2 ,
\label{eq:VH_universal}\\
V_{F_n} &= \frac{1}{2}L^2\rho^{4-n}\,\e^{a_n\phi}\,\widetilde F_n^2 .
\label{eq:VFn_universal}
\end{align}
The tilde quantities depend on flux quanta and on the non-universal internal fields, such as fields governing the shape of the internal manifold, but not on the universal fields $L$, $\rho$ or $\phi$. In particular, indices are raised and lowered with the tilde reference internal metric normalised to unit volume.

We consider then the contribution of localised $a=1,2,\dots$ sources wrapping a $2k_a$-dimensional cycle in the internal manifold and filling entirely the external 2D spacetime,
\begin{equation}
V_{loc,a} = T_{loc,a}\, L^2 \rho^{k_a} e^{\gamma_a \phi} ,
\end{equation}
such that the total 2D scalar potential is
\begin{equation}
V = V_R + V_H + \sum_n V_{F_n} +\sum_a V_{loc, a}\, .
\label{eq:V2Dgen}
\end{equation}
The coefficients $\gamma_a$ of the dilaton coupling depend on the nature of the source and will be specified later on; $T_{loc,a}$ contains the tension of the objects and can potentially depend on other fields.

Since there is no distinguished 2D Einstein frame, and since we assume constant scalars, the natural vacuum equations are simply the stationarity conditions of the reduced action with respect to universal fields,
\begin{equation}
L\partial_L\mathcal{L}=0,
\qquad
\rho\partial_\rho\mathcal{L}=0,
\qquad
\partial_\phi\mathcal{L}=0.
\label{eq:stationarity_universal}
\end{equation}
Using the scalings in \eqref{eq:VR_universal}--\eqref{eq:VFn_universal}, one finds
\begin{align}
L\partial_L\mathcal{L}=0 &:\qquad V_R=-V_H-\sum_n V_{F_n} - \sum_a V_{loc, a}\, ,
\label{eq:L_equation_universal}\\
\rho\partial_\rho\mathcal{L}=0 &:\qquad 8\lambda\rho^4=3V_R+V_H+\sum_n (4-n)V_{F_n}+ \sum_a k_a V_{loc, a}\, ,
\label{eq:rho_equation_universal}\\
\partial_\phi\mathcal{L}=0 &:\qquad V_H=\sum_n a_n V_{F_n} +\sum_a \gamma_a V_{loc, a}\, .
\label{eq:phi_equation_universal}
\end{align}
These replace the usual higher-dimensional requirement that one extremizes an Einstein-frame scalar potential.

\subsection{Propagating universal fields in 2D}
\label{subsec:universal_dof}

Even if the 10D dilaton supplies one scalar degree of freedom, in 2D and in the presence of a potential the propagating mode may be a mixture of volume and dilaton fluctuations. It is useful to set this 2D dilaton field apart by introducing the combination of $\rho$ and $\phi$ that defines the string frame internal volume,
\begin{equation}
e^{2y} = \rho^4 e^{2\phi}.
\end{equation}
The transformation
\begin{equation}
 \rho\longrightarrow e^\Delta\rho,
 \qquad
 \phi\longrightarrow\phi-2\Delta,
\label{eq:background_scaling_bulk}
\end{equation}
leaves $y$ invariant. For the supersymmetric Minkowski solutions considered below, it maps one solution into another and therefore parametrises a one-parameter family of backgrounds. As we will explain, the corresponding direction is non-propagating in two dimensions and should not be interpreted as an additional massless scalar. This is indeed related to the apparent presence of a ghost-like direction in the kinetic terms of $\rho$.

Retaining the spacetime dependence of the volume and 10D dilaton while suppressing any other possible matter fields gives, up to boundary terms,
\begin{equation}
S_{\rm univ}=2\pi\int \d^2x\sqrt{-\tilde g_2}\left[
\rho^4\tilde R_2+14\rho^4\bigl(\tilde\partial\log\rho\bigr)^2
-\frac{1}{2}\rho^4(\tilde\partial\phi)^2-V(\rho,\phi)\right].
\label{eq:universal_local_action}
\end{equation}
Thus, $\rho$ and $\phi$ are independent fields off shell. However, $\rho^4$, which measures the internal volume, is the 2D gravitational dilaton because it multiplies $\tilde R_2$. Its apparently wrong-sign kinetic term cannot be interpreted independently of the 2D metric equation of motion constraints.

To see this, one can take the variation of the above action with respect to the 2D metric $\tilde g_{\mu\nu}$ and linearize it around the supersymmetric Minkowski vacuum of interest, where all scalars are constant and $V=0$. The resulting equation for the fluctuation of $\rho^4$, which we call $\xi$ is 
\begin{equation}
\left(\eta_{\mu\nu} \widetilde{\vphantom{A}\Box}\ - \partial_\mu\partial_\nu \right)\xi=0.
\end{equation}
Taking the trace gives $\widetilde{\vphantom{A}\Box}\ \xi=0$ and thus we get the stronger condition
 \begin{equation}
 \partial_\mu \partial_\nu\xi=0.
 \end{equation}
 The general solution is $\xi=a + b_\mu x^\mu$, but for its interpretation as a small perturbation around a fixed static vacuum we should require $b_\mu=0$, otherwise $\xi$ would grow arbitrarily large for large times and distances (thus  jeopardizing normalisability). Then, $\xi=a$ is a constant whose kinetic term is vanishing and thus $\rho^4$ does not fluctuate. Hence, propagating fluctuations around the vacuum are such that $\delta y = \delta \phi$. The kinetic term \eqref{eq:universal_local_action} restricted along this subspace gives
 \begin{equation}
 \begin{aligned}
{\rm kin}|_{\delta y = \delta \phi} &=-\frac12 \rho^4 (\tilde \partial \delta \phi)^2,
\end{aligned}
 \end{equation}
which has the correct sign. This shows that no ghost is propagating in the model. Both dilaton and volume are independent in the reduced action, while only one combination carries a local degree of freedom. The orthogonal combination is a non-propagating mode or background parameter.

\subsection{No-go for vacua and scale separation}

Combining \eqref{eq:L_equation_universal}-\eqref{eq:phi_equation_universal} yields non-trivial information on the allowed vacua. First eliminate $V_R$ from the $\rho$-equation using the $L$-equation,
\begin{equation}
8\lambda\rho^4=-2V_H+\sum_n (1-n)V_{F_n} +\sum_a (k_a-3) V_{loc, a}.
\label{eq:pre_master_universal}
\end{equation}
Then, use the  $\phi$-equation together with $a_n=(5-n)/2$ to obtain
\begin{equation}
8\lambda\rho^4=-4\sum_n V_{F_n} +\sum_a(k_a-3-2\gamma_a) V_{loc, a}.
\label{eq:master_flux_identity_universal}
\end{equation}
Consider first the sourceless case, $V_{loc, a}=0$. 
If we turn on any RR fluxes, then $V_{F_n}$ is positive. This implies immediately that any sourceless critical point must satisfy
\begin{equation}
\lambda<0,
\label{eq:ads_only_universal}
\end{equation}
and $\lambda = 0$ only for the trivial case without flux. 
Hence, we find that sourceless tree-level type II compactifications with internal fluxes can only produce $\mathrm{AdS}_2$ vacua. 
This is the 2D version of the celebrated Maldacena--Nu\~nez logic forbidding sourceless Minkowski or de Sitter compactifications supported only by fluxes \cite{gibbons1985aspects,deWit:1986mwo, Maldacena:2000mw}.
It is useful to rewrite \eqref{eq:master_flux_identity_universal} as
\begin{equation}
V_{\lambda}\equiv -2\lambda\rho^4=\sum_n V_{F_n} -\frac{1}{4} \sum_a (k_a-3-2 \gamma_a)V_{loc,a}\, .
\label{eq:Vlambda_definition_universal}
\end{equation}
In the sourceless case, the entire negative cosmological constant is therefore set by positive flux energy.

The same equations \eqref{eq:L_equation_universal}-\eqref{eq:phi_equation_universal} constrain the possibility of achieving (parametric) scale separation on AdS$_2$ vacua. On the one hand, from the definitions of $V_R$ and $V_{\lambda}$, $i.e.$ \eqref{eq:VR_universal} and \eqref{eq:Vlambda_definition_universal}, and using that $\lambda=-1$ for $\mathrm{AdS}_2$ vacua, one has
\begin{equation}
-\frac{V_R}{V_{\lambda}}=\frac{L^2}{\rho}\,\frac{\widetilde R_8}{2} .
\label{eq:ratio_geometry_side_universal}
\end{equation}
As long as the normalised internal curvature $\widetilde R_8$ is of order one, parametric scale separation is the requirement
\begin{equation}
-\frac{V_R}{V_\lambda} \simeq \frac{L^2}{\rho} \gg 1\, .
\end{equation}
On the other hand, using the $\phi$-equation inside the $L$-equation gives
\begin{equation}
V_R=-\frac{1}{2}\sum_n (7-n)V_{F_n} -\sum_a(1+\gamma_a)V_{loc, a}.
\label{eq:VR_weighted_universal}
\end{equation}
Dividing by \eqref{eq:Vlambda_definition_universal} yields
\begin{equation}
-\frac{V_R}{V_\lambda} =  \frac{\frac{1}2 \sum_n  (7-n) V_{F_n}+\sum_a(1+\gamma_a)V_{loc,a}}{\sum_n V_{F_n}-\frac{1}4 \sum_a(k_a-3-2\gamma_a)V_{loc,a}}\,.
\label{eq:ratio_flux_side_universal}
\end{equation}
For the usual $Op/Dp$-brane sources filling the 2D spacetime, we have (in Einstein frame) $k_{p} = \frac{p-1}{2}$, $\gamma_{p} = \frac{p-3}{4}$ such that $k_p-3-2\gamma_p = -2$ and thus we get
\begin{equation}
\label{eq:VratioscalesepDpOp}
-\frac{V_R}{V_\lambda} =  \frac{\frac{1}2 \sum_n  (7-n) V_{F_n}+\frac{1}4\sum_a(p+1)V_{Dp/Op}}{\sum_n V_{F_n}+\frac{1}2 \sum_aV_{Dp/Op}}\,.
\end{equation}
If negative tension sources are absent, namely $V_{Op} =0$, and assuming again $\tilde R_8$ of order one,  then the above ratio is upper bounded by $7/2$ (for $p\leq 6$) and scale separation is obstructed,\footnote{For D7/O7 or D8/O8 the bound could be slightly larger but is still of order one.}
\begin{equation}
V_{Op} = 0 \qquad \Rightarrow \qquad -\frac{V_R}{V_\lambda} \leq \frac{7}{2} \qquad \Rightarrow \qquad \frac{L^2}{\rho} \lesssim \mathcal{O}(1).
\end{equation}
The subcase in which also $V_{Dp}=0$ leads to the same result.

\subsection{A way out: negative-tension sources}
\label{subsec:negative_tension_universal_equations}

As it happens for $D>2$ \cite{Gautason:2015tig, Junghans:2020acz, Tringas:2025uyg}, orientifolds provide a way out to the no-go theorem for scale separation.
Negative-tension objects correspond to $T_{\rm loc,a}<0$, hence $V_{\rm loc,a}<0$. 
From \eqref{eq:Vlambda_definition_universal} we see how they can evade the previous no-go. In the absence of sources, the right-hand side is positive, so only $\mathrm{AdS}_2$ is possible. We have already discussed that $Dp$-branes contribute positively due to their universal coefficient $k_a-3-2\gamma_a=-2$ and to $T_{\rm loc, a}>0$; hence, they just add up to fluxes. 
Instead, once negative-tension sources are included, their contributions can compensate the positive flux energies and allow Minkowski or, in broader settings, even de Sitter solutions~\cite{Maldacena:2000mw, Giddings:2001yu, Hertzberg:2007wc, VanRiet:2023pnx}. 

Typical examples are the standard orientifold $Op$-planes for which in Einstein frame one has
\begin{equation}
V_{Op}= -|T_{Op}| L^2\rho^{\frac{p-1}{2}}\e^{\frac{p-3}{4}\phi},
\qquad p=\left\{\begin{array}{ll}
\text{even} & \text{in type IIA},\\[2pt]
\text{odd} & \text{in type IIB},
\end{array}\right.
\label{eq:Op_scaling_2d}
\end{equation}
so that $k_{Op}=(p-1)/2$ and $\gamma_{Op}=(p-3)/4$, as we have seen before. 
Since $k_p-3-2\gamma_p=-2$ and $V_{Op}<0$, orientifolds can evade the bound because the denominator of \eqref{eq:VratioscalesepDpOp} is no longer a positive sum, weakening the Maldacena--Nu\~nez obstruction. However, we have argued in the previous section that standard orientifolds $Op$ planes are not compatible with our requirement of preserving minimal SUSY in 2D after compactification on a $\mathrm{Spin}(7)$ manifold. Therefore, we introduced the less standard $\mathrm{OF1}$ and $\mathrm{ON5}$ planes \cite{Hanany:2000fq}. As we explained in section \ref{sec:commentsorientifolds}, these have different dilaton weights than \eqref{eq:Op_scaling_2d}.

In the quotient space, a single codimension-eight defect $\mathrm{OF1}$ has one sixteenth of the tension and charge of a fundamental
string, with opposite signs \cite{Hanany:2000fq,Dasgupta:1997cd}. Thus its contribution to the scalar potential scales as
\begin{equation}
V_{\mathrm{OF1}} = -2^{-4} |T_{F1}|L^2 \e^{\phi/2},
\label{eq:OF1_scaling}
\end{equation}
such that $k_{\mathrm{OF1}}=0$ and $\gamma_{\mathrm{OF1}}=1/2$, while an $\mathrm{ON5}$ source has the tension scaling of an NS5-brane,
\begin{equation}
V_{\mathrm{ON5}}= - |T_{NS5}| L^2\rho^2\e^{-\phi/2},
\label{eq:ON5_scaling}
\end{equation}
such  that $k_{\mathrm{ON5}}=2$ and $\gamma_{\mathrm{ON5}}=-1/2$.
Inserting these weights into \eqref{eq:ratio_flux_side_universal} shows that $\mathrm{OF1}$ contributes with coefficient $+1$ in the denominator, whereas the contribution from an $\mathrm{ON5}$ drops out of that particular linear combination in the denominator altogether. This does not make $\mathrm{ON5}$ irrelevant: it still appears in the separate $L$-, $\rho$- and $\phi$-equations, and precisely because its dilaton scaling differs from that of the RR sector it is a plausible ingredient for lifting universal flat directions. For the next section, however, we will focus on the simpler setup in which $\mathrm{OF1}$s are the only localised sources.

\section{The 2D theory from a toroidal $\mathrm{Spin}(7)$ orbifold}
\label{sec:toroidal_bulk_moduli}

Minimal supersymmetry in 2D naturally leads one to 8D spaces with $\mathrm{Spin}(7)$-holonomy. Compact $\mathrm{Spin}(7)$ manifolds provide Ricci-flat internal spaces with the smallest amount of supersymmetry compatible with a geometric compactification, while their calibrated forms furnish a natural language for supersymmetry conditions and flux couplings \cite{HarveyLawson1982, JoyceBook2000, Becker:2000jc, Acharya:2002vs, Becker:2003wb, Bonetti:2013nka}.\footnote{In the absence of fluxes and localised sources, supersymmetry can also be maintained after including $\alpha'$ corrections, provided the internal metric and supersymmetry transformations are corrected appropriately \cite{Becker:2014rea}.} Compact $\mathrm{Spin}(7)$ manifolds also admit type IIA orientifolds that introduce negative-tension objects needed for Minkowski vacua. Rather than attempting an exhaustive classification, we isolate a deliberately simple corner in which the universal equations can be derived explicitly, the role of negative-tension objects can be analysed directly, and the bulk cohomology of a toroidal $\mathrm{Spin}(7)$ model makes it possible to track fluxes and supersymmetry conditions analytically. This gives a controlled setting in which to study flux compactifications of string theory down to 2D.

We now construct the 2D effective theory of type II string theory compactified on a toroidal orbifold limit of a $\mathrm{Spin}(7)$-holonomy space. We include fluxes and localised sources, and analyse the resulting scalar fields and effective potential. In the untwisted bulk sector, the only non-vanishing Betti numbers are $b_0$, $b_4$, and $b_8$. This naturally leads us to work in type IIA, where RR fluxes can thread internal cycles of even degree.
The sparse untwisted cohomology makes the flux sector very constrained. We first treat the flux coefficients as continuous parameters while deriving the scalar potential and its $\mathcal N=(1,1)$ supergravity completion. Then, we impose flux quantisation on the resulting candidate vacua.

\subsection{Orbifold action, fluxes and sources}
Our example is based on Joyce's construction of compact $\mathrm{Spin}(7)$ spaces that arise from $\mathbb{T}^8/\mathbb{Z}_2^4$ toroidal orbifolds \cite{Joyce1996, JoyceBook2000}. We take the coordinates on $\mathbb{T}^8$ to have unit periodicity,
$x^i\sim x^i+1$, and quotient by the affine action
\begin{equation}
\Gamma=\langle \alpha,\beta,\gamma,\delta\rangle\simeq \mathbb Z_2^4 ,
\end{equation}
whose generators act as 
\begin{align}
\label{eq:joyce_orbifold_action1}
\alpha:\quad
(x^1,\ldots,x^8)
&\mapsto
(-x^1,-x^2,-x^3,-x^4,x^5,x^6,x^7,x^8),\\
\beta:\quad
(x^1,\ldots,x^8)
&\mapsto
(x^1,x^2,x^3,x^4,-x^5,-x^6,-x^7,-x^8),\\
\gamma:\quad
(x^1,\ldots,x^8)
&\mapsto
\left(-x^1,-x^2,x^3,x^4,-x^5,-x^6,x^7,x^8\right),\\
\delta:\quad
(x^1,\ldots,x^8)
&\mapsto
\left(\frac{1}{2}-x^1,x^2,\frac{1}{2}-x^3,x^4,
\frac{1}{2}-x^5,x^6,\frac{1}{2}-x^7,x^8\right),
\label{eq:joyce_orbifold_action4}
\end{align}
where all entries are understood modulo one. This orbifold is taken from section 14.5 of \cite{JoyceBook2000}. The half-shifts affect the fixed-point structure but drop out when the action is differentiated; the invariant constant forms therefore follow directly from \eqref{eq:joyce_orbifold_action1}--\eqref{eq:joyce_orbifold_action4}.  On the covering $\mathbb{T}^8$ these forms are harmonic, and the  $\Gamma$-invariant subset gives the untwisted harmonic forms of $\mathbb{T}^8/\Gamma$. It is generated by the scalar, the volume form,
\begin{equation}
{\rm dvol}_8 = {\rm dx}^{12345678}, \qquad \int_{\mathbb{T}^8} {\rm dvol}_8 = 1,
\end{equation}
and fourteen harmonic 4-forms, 
\begin{equation}
\begin{aligned}
\omega_1&=-\d x^{1234}, & \widetilde\omega_1&=-\d x^{5678}, \\
\omega_2&=-\d x^{1256}, & \widetilde\omega_2&=-\d x^{3478}, \\
\omega_3&=-\d x^{1278}, & \widetilde\omega_3&=-\d x^{3456}, \\
\omega_4&=-\d x^{1357}, & \widetilde\omega_4&=-\d x^{2468}, \\
\omega_5&=\d x^{1368}, & \widetilde\omega_5&=\d x^{2457}, \\
\omega_6&=\d x^{1458}, & \widetilde\omega_6&=\d x^{2367}, \\
\omega_7&=\d x^{1467}, & \widetilde\omega_7&=\d x^{2358}.
\label{eq:paired_bulk_forms}
\end{aligned}
\end{equation}
One can also check explicitly that this 4-form basis is such that
\begin{equation}
\label{eq:wtw=1}
\int_{\mathbb{T}^8} \omega_A \wedge \widetilde \omega_B = \delta_{AB}.
\end{equation}
We thus have the untwisted non-vanishing Betti numbers
\begin{equation}
 b_0^{\rm untw}=1,
 \qquad
 b_4^{\rm untw}=14,
 \qquad
 b_8^{\rm untw}=1\,,
\label{eq:untwisted_betti_numbers}
\end{equation}
with no invariant 1-, 2-, 3-, 5-, 6- or 7-forms. 

In the untwisted diagonal truncation, the bulk metric contains one field governing the overall volume and seven independent shape fields, in agreement with the general deformation theory of compact $\mathrm{Spin}(7)$ spaces \cite{Becker:2000jc, Acharya:2002vs}.
The reduction is performed on a diagonal metric Ansatz,
\begin{equation}
\d s_8^2=\rho\sum_{i=1}^8 e^{2\sigma_i}(\d x^i)^2,
\qquad \sum_{i=1}^8\sigma_i=0,
\label{eq:diag_metric_bulk}
\end{equation}
where $\rho$ is the internal volume modulus introduced in section \ref{sec:universal_constraints} and the seven independent combinations of the $\sigma_i$ describe the bulk shape deformations. 
The orbifold action also projects out every off-diagonal metric component.
The constraint $\sum_{i=1}^8 \sigma_i=0$ fixes the appropriate gauge in which $\rho$ is removed from the overall combination of the $\sigma_i$, which are thus pure shape moduli, namely  
\begin{equation}
{\rm Vol}(\mathbb{T}^8) =\int_{\mathbb{T}^8} {\rm d}^8x\sqrt{g_8} = \rho^4 e^{\sum_i \sigma_i} \int_{\mathbb{T}^8} {\rm d}^8x = \rho^4.
\end{equation}
Equivalently, this constraint removes the redundancy of a common shift $\sigma_i \to \sigma_i +\Delta$ that could be reabsorbed into $\rho \to \rho e^{-2\Delta}$. 

The signs in the 4-form basis \eqref{eq:paired_bulk_forms} are chosen so that at the isotropic point, where all $\sigma_i=0$, we have  $\star_8\,\omega_A=\widetilde\omega_A$. Introducing the seven independent shape moduli coordinates,
\begin{equation}
\begin{aligned}
X_1&=\sigma_1+\sigma_2+\sigma_3+\sigma_4, &
X_2&=\sigma_1+\sigma_2+\sigma_5+\sigma_6, &
X_3&=\sigma_1+\sigma_2+\sigma_7+\sigma_8, \\
X_4&=\sigma_1+\sigma_3+\sigma_5+\sigma_7, &
X_5&=\sigma_1+\sigma_3+\sigma_6+\sigma_8, &
X_6&=\sigma_1+\sigma_4+\sigma_5+\sigma_8, \\
X_7&=\sigma_1+\sigma_4+\sigma_6+\sigma_7,
\label{eq:Xvariables_bulk}
\end{aligned}
\end{equation}
and using $\sum_{i=1}^8\sigma_i=0$, we obtain
\begin{equation}
\star_8\,\omega_A=e^{-2X_A}\widetilde\omega_A,
\qquad
\star_8\,\widetilde\omega_A=e^{2X_A}\omega_A.
\label{eq:hodge_star_pairs}
\end{equation}
The Cayley form associated to the unit-volume metric is then 
\begin{equation}
\Phi_4=\sum_{A=1}^7\Bigl(e^{X_A}\omega_A+e^{-X_A}\widetilde\omega_A\Bigr),
\label{eq:normalized_cayley_bulk}
\end{equation}
and is such that
\begin{align}
\star_8 \Phi_4 = \Phi_4, \qquad \Phi_4 \wedge \star_8 \Phi_4 = 14\, {\rm dvol}_8.
\end{align}
At the isotropic point all $X_A=0$ and this reduces to the standard invariant Cayley calibration on the orbifold limit \cite{Joyce1999, HarveyLawson1982,  Becker:2000jc}, as in \eqref{eq:Phi4ei}. 

Given the above topology, when dualising all fluxes so that they are purely internal, the only RR fluxes that we can have are
\begin{equation}
F_0,
\qquad F_4,
\qquad F_8.
\label{eq:bulk_iia_fluxes}
\end{equation}
We cannot include ordinary spacetime-filling RR orientifold planes of type IIA in the untwisted bulk truncation, since they would wrap odd-dimensional internal cycles, whereas the untwisted invariant odd cohomology is absent. The natural negative-tension object is instead the spacetime-filling $\mathrm{OF1}$ \cite{Hanany:2000fq, Bergman:2001rp}, which does not wrap an internal cycle and carries the NSNS charge appearing in the magnetic Bianchi identity \cite{Cremmer:1998px, Banerjee:2026ptm}
\begin{equation}
\d H_7 =F_0\wedge F_8+\frac{1}{2} F_4\wedge F_4+Q_{\mathrm{OF1}}\,\delta_8.
\label{eq:H7_bianchi_bulk}
\end{equation}
This involves precisely the RR fluxes that survive in the bulk truncation.

Ideally we would also like to include $\mathrm{ON5}$ planes that wrap the internal 4-cycles that exist in our setup. The corresponding Bianchi identity is
\begin{equation}
\d H_3 = Q_{\mathrm{ON5}}\,\delta_4.
\label{eq:H3_bianchi}
\end{equation}
Integrating this equation over compact 4-cycles implies that the total $\mathrm{ON5}$ charge in each cohomology class must vanish. Thus $\mathrm{ON5}$ planes cannot be included with uncancelled net charge. They would have to be accompanied by NS5 branes, or arranged so that the total source is cohomologically trivial. In the smeared approximation relevant for the present bulk analysis, such locally cancelled $\mathrm{ON5}$/NS5 systems do not generate an independent net source term, and we will not include them.

\subsection{A smooth resolution and the normalisation of bulk forms}
\label{sec:smoothres}

From Lemma 14.5.1 of \cite{JoyceBook2000}, we learn that the orbifold under consideration has several singular loci. The fixed loci of $\alpha$, $\beta$, $\gamma$, $\alpha \gamma$, $\beta \gamma$, $\alpha \beta \gamma$ and $\delta$ are each 16 copies of $\mathbb{T}^4 \subset \mathbb{T}^8$, while $\alpha \beta$ has 256 fixed points on $\mathbb{T}^8$. 
As explained in  \cite{JoyceBook2000}, their resolution can conveniently be organized in three steps.  

One first resolves $\mathbb T^8/\langle\alpha,\beta\rangle$ to $K3_\alpha\times K3_\beta$ and then lifts the residual $\langle\gamma,\delta\rangle$ action in two steps.  The integers $k,l\in\{0,\ldots,8\}$ count, on the two $K3$ factors, the pairs of singular points for which one chooses one of the two possible lifts of $\gamma$. The additional integers $a,c$ encode the numbers of connected components of the fixed surfaces, and the integers $b,d$ the subset of these pairs of connected components exchanged by $\delta$. In particular, $2b \leq a$ and $2d \leq c$.  According to Theorem~14.5.4 of \cite{JoyceBook2000}, the resulting smooth compact manifold $X_8$ has then the Betti numbers
\begin{align}
 b_2&=2+k+l+bc+ad-2bd,\\
 b_3&=8+8a+8c-2al-2kc+2ac,\\
 b_4^+&=237-33k-33l+8kl+8a+8c-2al-2kc+2ac-bc-ad+2bd,\\
 b_4^-&=109-17k-17l+4kl+8a+8c-2al-2kc+2ac-bc-ad+2bd,\\
 b_4&=346-50k-50l+12kl+16a+16c-4al-4kc+4ac
      -2bc-2ad+4bd,
\label{eq:third_joyce_betti_general}
\end{align}
where $b_4 = b_4^+ + b_4^-$.
Substitution into $\chi=2+2b_2-2b_3+b_4$ gives the particularly simple result
\begin{equation}
 \chi(X_8)=336-48k-48l+12kl,
 \qquad k,l=0,\ldots,8,
\label{eq:third_joyce_chi_general}
\end{equation}
independently of $a,b,c,d$.

For the choice $k=l=0$, $a=c=1$, $b=d=0$, which is one of the allowed set of values \cite{JoyceBook2000}, one obtains
\begin{equation}
 b_2=2,
 \qquad
 b_3=26,
 \qquad
 b_4^+=255,
 \qquad
 b_4^-=127,
 \qquad
 b_4=382,
\label{eq:third_joyce_betti_k0}
\end{equation}
and hence
\begin{equation}
 \chi(X_8)=2+2b_2-2b_3+b_4=336\,.
\label{eq:third_joyce_chi_336}
\end{equation}
This is the first resolution in Table 14.2 of \cite{JoyceBook2000}. In the following, its topological data will refer to the smooth manifold $X_8$, while all explicit calculations involving the coordinate forms and the diagonal toroidal metric will be performed at the orbifold point $\mathbb T^8/\Gamma$, within the untwisted bulk approximation.

At the orbifold point, the covering-space forms are not conveniently normalised: since $\Gamma$ has order sixteen, we have
\begin{equation}
\int_{\mathbb T^8/\Gamma} {\rm d}x^{12345678} = \frac{1}{16}\,,\qquad
\int_{\mathbb T^8/\Gamma} \omega_A\wedge\widetilde\omega_B = \frac{1}{16}\,\delta_{AB}\,.
\end{equation}
To keep the effective action of section~\ref{subsec:reduced_potential} free of such factors, we absorb them in a formal rescaling $dx^i \to \sqrt{2} dx^i$. We will use the same notation ${\rm dvol}_8$, $\omega_A$ and $\widetilde\omega_A$ to denote the rescaled invariant forms
\begin{equation}
{\rm dvol}_8 = 16\,{\rm d}x^{12345678}\,,\qquad
\omega_A \to 4\,\omega_A\,,\qquad
\widetilde\omega_A \to 4\,\widetilde\omega_A\,,
\label{eq:working_normalisation}
\end{equation}
so that on the quotient
\begin{equation}
\int_{\mathbb T^8/\Gamma} {\rm dvol}_8 = 1\,,\qquad
\int_{\mathbb T^8/\Gamma} \omega_A\wedge\widetilde\omega_B = \delta_{AB}\,,
\label{eq:quotient_unit_pairing}
\end{equation}
while the Hodge relations \eqref{eq:hodge_star_pairs} and the Cayley form \eqref{eq:normalized_cayley_bulk} keep their form. We stress that this is just a matter of conventions: flux quantisation, to which we now turn, must be imposed separately.

Especially in the presence of Romans mass, RR flux quantisation is a subtle problem. At the level of supergravity, one usually imposes Page flux quantisation \cite{Marolf:2000cb}, which in our conventions ($2\pi\sqrt{\alpha'}=1$), and in the gauge $B_2=0$ used here, amounts to requiring $\int_C F_4\in\mathbb Z$ for every integral 4-cycle $C\subset X_8$, i.e.\ $[F_4]\in H^4(X_8,\mathbb Z)$. 
For a purely untwisted flux, we derive candidate Page quantisation conditions from the fourteen explicit orbifold coordinate cycles constructed below. Interpreting these conditions as Page quantisation conditions on the smooth resolution requires the additional assumption that these orbifold cycles lift to integral 4-cycles of $X_8$ and that the periods of the untwisted forms are unchanged under this lift. Nevertheless, more refined RR quantisation conditions have been formulated \cite{Diaconescu:2000wy}. Since we are unable to implement them explicitly for our toroidal example, we restrict ourselves to Page flux quantisation. The obstruction found below is therefore conditional both on this quantisation prescription and on the cycle-lift assumption, and could be modified or even lifted if either assumption is changed.

We now determine the periods of the fourteen invariant 4-forms \eqref{eq:paired_bulk_forms} over a convenient set of cycles at the orbifold point. For the time being, ${\rm d}x^i$ denotes the original unit-period differential on the covering torus, with $x^i\sim x^i+1$, rather than the rescaled one introduced in \eqref{eq:working_normalisation}. We label each of these fourteen 4-forms by a multi-index
$S=(i_1i_2i_3i_4)$ with $i_1<i_2<i_3<i_4$. Then, we set ${\rm d}x^S\equiv {\rm d}x^{i_1}\wedge{\rm d}x^{i_2}\wedge{\rm d}x^{i_3}\wedge{\rm d}x^{i_4}$, and take a slice $\Sigma_S =\{x^j = c^j\,\,\, \text{for} \,\, j \notin S\}\cong \mathbb{T}^4$ tangent to $S$, with $c^j$ constants; we denote its image in $\mathbb T^8/\Gamma$ by $C_S$.
Let $H_S\subset\Gamma$ be the subgroup preserving $\Sigma_S$, and let $K_S\subset H_S$ be the kernel of its action on $\Sigma_S$, namely the subgroup acting trivially on every point of the slice. The group acting effectively on $\Sigma_S$ is therefore $H_S/K_S$, with order
\begin{equation}
k_S=\left|H_S/K_S\right|.
\end{equation}
Away from its fixed loci, the quotient map $\Sigma_S\rightarrow C_S$ is $k_S$-to-one. Since $\int_{\Sigma_S}{\rm d}x^S=1$, the period of the descended coordinate form over the resulting closed orbifold 4-cycle is
\begin{equation}
\left|\int_{C_S}{\rm d}x^S\right|=\frac{1}{k_S}.
\end{equation}
The values of $k_S$ follow directly from \eqref{eq:joyce_orbifold_action1}--\eqref{eq:joyce_orbifold_action4}. We choose the transverse constants $c^j$ in $\{0,\tfrac12\}$. Since $-c^j=c^j$ modulo one for these values, every such slice is preserved by the sign subgroup $\Gamma_0=\langle\alpha,\beta,\gamma\rangle\simeq\mathbb Z_2^3$, which has order 8. However, an element of $\Gamma_0$ that reflects only transverse directions acts trivially on $\Sigma_S$ and therefore belongs to $K_S$. The sets of coordinate directions reflected by the seven nontrivial elements of $\Gamma_0$ are
\begin{equation}
1234,\qquad 5678,\qquad 1256,\qquad 3478,\qquad
1278,\qquad 3456,\qquad 12345678.
\end{equation}
For each slice tangent to one of the first six sets, exactly one nontrivial element of $\Gamma_0$ reflects only the complementary transverse directions. These elements are, respectively, $\beta$, $\alpha$, $\alpha\beta\gamma$, $\gamma$, $\alpha\gamma$, and $\beta\gamma$. Consequently, the kernel of the action of $\Gamma_0$ has order two, so that $\Gamma_0$ acts effectively through a group of order $8/2=4$.
For the remaining eight invariant directions, no nontrivial element of $\Gamma_0$ acts trivially on the slice, so $\Gamma_0$ acts effectively through a group of order eight.
It remains to consider the eight elements of $\Gamma$ that do not belong to $\Gamma_0$. Each of them is obtained by composing $\delta$ with an element of $\Gamma_0$. On the chosen slices, such an element preserves $\Sigma_S$ only when all four directions $1357$ are tangent to the slice. Therefore, for every $S\neq1357$, one has $H_S=\Gamma_0$, giving $k_S=4$ for the first six directions and $k_S=8$ for the remaining seven. For $S=1357$, the full group $\Gamma$ preserves the slice and acts effectively on it, giving
\begin{equation}
k_{1357}=16.
\end{equation} 
We thus obtain the following periods for the chosen coordinate cycles:
\begin{center}
\begin{tabular}{c|c|c}
directions $S$ & pair label $A$ & orbifold period $1/k_S$\\
\hline
$1234, 5678, 1256, 3478, 1278, 3456$ & $1,2,3$ & $1/4$\\
$1368, 2457, 1458, 2367, 1467, 2358$ & $5,6,7$ & $1/8$\\
$1357$ & $4$ & $1/16$\\
$2468$ & $4$ & $1/8$
\end{tabular}
\end{center}
Here $A$ labels the pairs $(\omega_A,\widetilde\omega_A)$ defined in \eqref{eq:paired_bulk_forms}. These are periods over explicit orbifold coordinate cycles, but they might not be the smallest periods over the full integral homology lattice. To use them as Page periods on the smooth Joyce manifold $X_8$, we assume that the corresponding orbifold cycles lift to integral 4-cycles of the chosen resolution and that the periods of the untwisted forms are unchanged under this lift. Under this assumption, the cycles above provide necessary Page flux quantisation conditions.

Going back to the normalisation \eqref{eq:working_normalisation} we have $\omega_A=\pm 4\,{\rm d}x^{S_A}$ and $\widetilde\omega_A=\pm 4\,{\rm d}x^{\widetilde S_A}$, so the periods of a purely untwisted flux
\begin{equation}
F_4=\sum_{A=1}^7\left(N_A\omega_A+M_A\widetilde\omega_A\right)
\end{equation}
over these cycles are $4N_A/k_{S_A}$ and $4M_A/k_{\widetilde S_A}$, up to signs. Integrality on all fourteen cycles then requires
\begin{equation}
N_{1,2,3},\,M_{1,2,3}\in\mathbb Z\,,\qquad
N_{5,6,7},\,M_{5,6,7}\in2\mathbb Z\,,\qquad
N_4\in4\mathbb Z\,,\qquad
M_4\in2\mathbb Z\,.
\label{eq:corrected_quantisation}
\end{equation}
Under the cycle-lift assumption stated above, these are necessary Page flux quantisation conditions and are sufficient to derive the conditional obstruction below.

 \subsection{Dimensional reduction to 2D}
\label{subsec:reduced_potential}

We now derive the 2D effective theory along the lines of section \ref{subsec:reduced_action_2d}. 
To perform the dimensional reduction, we make the flux Ansatz
\begin{equation}
F_0=m,
\qquad
F_4=\sum_{A=1}^7\Bigl(N_A\,\omega_A+M_A\,\widetilde\omega_A\Bigr),
\qquad
F_8= n_8\,\mathrm{dvol}_8.
\label{eq:flux_ansatz_bulk_section}
\end{equation}
In our conventions with $2\pi \sqrt{\alpha'}=1$, flux quantisation requires $m,n_8\in\mathbb Z$ together with integrality of the class $[F_4]$ as discussed above; see \eqref{eq:corrected_quantisation}. For the dimensional reduction we keep $N_A$ and $M_A$ continuous and impose these quantisation conditions only when analysing the candidate vacua. As explained in the previous section, the appropriate normalisation of the bulk forms is understood after descent to the orbifold; we keep denoting $\d x^i$, $\omega_A$ and $\widetilde\omega_A$  the quotient-normalised invariant forms, such that $\int_{\mathbb{T}^8/\Gamma} \mathrm{dvol}_8=1$, $ \int_{\mathbb{T}^8/\Gamma} \omega_A\wedge\widetilde\omega_B = \delta_{AB}$ and also $ds_8^2 = 2 \rho \sum_{i=1}^8 e^{2\sigma_i}(dx^i)^2$.
Using \eqref{eq:hodge_star_pairs} one finds then
\begin{align}
\int_{\mathbb{T}^8/\Gamma}F_4\wedge *_8F_4
&=\sum_{A=1}^7\left(N_A^2e^{-2X_A}+M_A^2e^{2X_A}\right),
\label{eq:F4_norm_bulk_section}\\
\int_{\mathbb{T}^8/\Gamma}F_8\wedge *_8F_8
&=n_8^2\rho^{-4},
\label{eq:F8_norm_bulk_section}\\
\int_{\mathbb{T}^8/\Gamma}F_0\wedge *_8F_0
&=m^2\rho^{4}.
\label{eq:F0_norm_bulk_section}
\end{align}
The explicit dimensional reduction is performed using the flat diagonal metric at the orbifold point $\mathbb T^8/\Gamma$. This metric is Ricci-flat in the orbifold sense, so that $\widetilde R_8=0$ and hence $V_R=0$ in \eqref{eq:VR_universal}.\footnote{The smooth resolution $X_8$ also admits a torsion-free $\mathrm{Spin}(7)$ structure by Theorem 13.6.2 of \cite{JoyceBook2000}, but its explicit Ricci-flat metric is not the diagonal toroidal metric used in the calculation.} Since we do not have $H_3$ flux either, it only remains to include the contribution from the $\mathrm{OF1}$ sources \eqref{eq:OF1_scaling}, which we write as
\begin{equation}
V_{\mathrm{OF1}}=-T_{\mathrm{OF1}} L^2 e^{\phi/2},
\qquad T_{\mathrm{OF1}} > 0.
\label{eq:OF1_potential_bulk_section}
\end{equation}
Using \eqref{eq:VR_universal}-\eqref{eq:VFn_universal} we obtain the 2D Lagrangian 
\begin{align}
\mathcal{L}=& 2\lambda\rho^4 + {\rm kin}-
\frac{L^2e^{\phi/2}}{2}\sum_{A=1}^7\left(N_A^2e^{-2X_A}+M_A^2e^{2X_A}\right)\cr
&-\frac{L^2}{2}m^2\rho^4e^{5\phi/2}
-\frac{L^2}{2}n_8^2\rho^{-4}e^{-3\phi/2}
+T_{\mathrm{OF1}} L^2e^{\phi/2},
\label{eq:full_bulk_potential_section}
\end{align}
where ${\rm kin}$ contains the kinetic terms of $\rho$, $\phi$ and $X_A$, which we will calculate shortly. 
This is the explicit toroidal realization of the effective theory \eqref{eq:S2_schematic_universal}.

The 2D kinetic term for the dilaton stems directly from dimensional reduction of the 10D one and it thus reads $-\frac{1}{2} \rho^4 (\tilde \partial\phi)^2$, where we have inserted already the factor $\sqrt{g_8}= \rho^4$. The kinetic terms of $\rho$ and $X_A$, instead, arise from dimensional reduction of the 10D Ricci scalar, $R_{10}$. One can calculate 
\begin{equation}
L^2R_{10 } =\tilde R_2 -8\widetilde{\vphantom{A}\Box}\log\rho -18(\tilde \partial\log\rho)^2  -\sum_{i=1}^8(\tilde\partial\sigma_i)^2,
\end{equation}
with $\tilde R_2=L^2 R_2= 2\lambda$ giving the first term on the right hand side of \eqref{eq:full_bulk_potential_section}. Tildes denote quantities computed using the normalized external metric in \eqref{eq:metric_ansatz_universal}.
Notice that, due to linearity of the definitions of $X_A$ in \eqref{eq:Xvariables_bulk}, and using $\sum_{i=1}^8 \sigma_i=0$, we can write
\begin{equation}
\sum_{i=1}^8 (\tilde \partial \sigma_i)^2 = \frac{1}{2} \sum_{A=1}^7 (\tilde\partial X_A)^2,
\end{equation}
and thus, inserting again $\sqrt{g_8}= \rho^4$,
\begin{equation}
\begin{aligned}
{\rm kin} &= -8\rho^4\, \widetilde{\vphantom{A}\Box}\log\rho -18\rho^4(\tilde \partial\log\rho)^2  -\rho^4\sum_{i=1}^8(\tilde\partial\sigma_i)^2-\frac{1}{2} \rho^4 (\tilde \partial\phi)^2\\
&=14 \rho^4(\tilde \partial\log\rho)^2  -\frac{1}{2}\rho^4\sum_{A=1}^7(\tilde\partial X_A)^2-\frac{1}{2} \rho^4 (\tilde \partial\phi)^2,
\end{aligned}
\end{equation}
where in the second step we neglected total derivatives.
Next, introducing the following combination of dilaton and volume
\begin{equation}
e^{2y} = \rho^4 e^{2\phi},
\end{equation}
implying $\partial \log \rho = \frac{1}{2}(\partial y-\partial \phi)$, and grouping the scalars into $q^I = (X_A, y, \phi)$, we can write all kinetic terms collectively as
\begin{equation}
{\rm kin} = -\frac{\rho^4}{2}\mathcal{G}_{IJ}\tilde \partial_\mu q^I \tilde \partial^\mu q^J,
\end{equation}
with
\begin{equation}
\mathcal{G}_{IJ} = \left(
\begin{array}{ccc}
\delta_{AB} & 0 &0\\
0 &-7&7\\
0&7&-6
\end{array}
\right) \qquad {\rm and} \qquad \mathcal{G}^{IJ} = \left(
\begin{array}{ccc}
\delta^{AB} & 0 &0\\
0 &\frac{6}{7}&1\\
0&1&1
\end{array}
\right).
\end{equation}
In the new variables, the scalar potential reads
\begin{equation}
V=\frac{1}{2}L^2 e^{\phi/2}\left[
\sum_{A=1}^7\left(N_A^2e^{-2X_A}+M_A^2e^{2X_A}\right)
+ m^2 e^{2y}
+n_8^2 e^{-2y}
-2T_{\mathrm{OF1}}\right]\,.
\label{eq:V_flux_source_y_bulk}
\end{equation}
and the complete 2D effective action  is finally
\begin{equation}
S =2\pi \int d^2x \sqrt{-\tilde g_2} \left(2\lambda\rho^4- \frac{\rho^4}{2}\mathcal{G}_{IJ}\tilde \partial_\mu q^I \tilde \partial^\mu q^J-V\right),
\label{eq:S2Ddimred}
\end{equation}
with $\tilde R_2 = 2 \lambda$.
The negative determinant ${\rm det}\mathcal{G}_{IJ}=-7$ does not signal a physical ghost. As explained in section~\ref{subsec:universal_dof}, $\rho^4$ is the gravitational dilaton, and the metric constraints remove its fluctuation around a Minkowski vacuum. One propagating universal scalar combination remains, with a positive kinetic term on the constrained fluctuation space. 

The parameter $T_{\mathrm{OF1}}>0$ is fixed in our compact model. As explained below, the $\mathrm{OF1}$ planes are the local description of the familiar 2D $B_2$-tadpole \cite{Sethi:1996es, Gukov:2002iq}. In our conventions, the magnetic equation for $B_2$ gives the integrated condition 
\begin{equation}
0=\int_{8} d H_7
=\int_{8}\left(F_0\wedge F_8+\frac{1}{2} F_4\wedge F_4\right)+Q_{\mathrm{OF1}},
\label{eq:integrated_B2_tadpole_bulk}
\end{equation}
and the total negative fundamental-string charge associated with the $\mathrm{OF1}$ orbifold system is\footnote{Here $\chi$ denotes the supersymmetric orbifold index entering the two-dimensional tadpole. From \eqref{eq:joyce_orbifold_action1}--\eqref{eq:joyce_orbifold_action4},
the element $\alpha\beta$ has $256$ internal fixed points on the covering $\mathbb T^8$, which are paired by $\delta$ into $128$ inequivalent quotient loci. Each locus has local stabilizer $\langle\alpha,\beta,\gamma\rangle\simeq\mathbb Z_2^3$, so these are not isolated $\mathbb R^8/\mathbb Z_2$ $\mathrm{OF1}$ singularities.
We therefore use the full orbifold index rather than assigning the isolated charge $-1/16$ to each quotient locus or dividing the covering-space count naively by $|\Gamma|=16$.}
\begin{equation}
Q_{\mathrm{OF1}}=-\frac{\chi}{24}.
\label{eq:QOF1_chi_bulk}
\end{equation}
For the orbifold CFT under consideration, $\chi=336$, and thus
\begin{equation}
\frac{\chi}{24}=14,\qquad
Q_{\mathrm{OF1}}=-14,\qquad
T_{\mathrm{OF1}}=14.
\label{eq:TOF1_six_bulk}
\end{equation}
Here, we used that the full localized orbifold system is BPS and
satisfies $T_{\mathrm{OF1}}=|Q_{\mathrm{OF1}}|$. Employing the flux Ansatz \eqref{eq:flux_ansatz_bulk_section}, the compact tadpole condition becomes
\begin{equation}
 m\, n_8+\sum_{A=1}^7 N_AM_A =T_{\mathrm{OF1}}=14.
\label{eq:tadpole_six_bulk}
\end{equation}
Notice that it is the signed bilinear $N_AM_A$, not its absolute value, that enters the constraint.

The $\mathrm{OF1}$ system appearing at the orbifold point should not be regarded as a leading-order effect of classical two-derivative type IIA supergravity. On the smooth resolution $X_8$, the corresponding fundamental string charge is generated by the one-loop higher-derivative coupling $B_2\wedge \mathcal{X}_8(R)$, whose integral gives $-\chi(X_8)/24$; the associated contribution to the energy arises from the $\mathcal{E}_8(R)$ term in its supersymmetric completion \cite{Sethi:1996es,Gukov:2002iq}.
In the orbifold limit, the cycles introduced in resolving the singularities shrink to zero size, and the curvature becomes string-scale in their neighborhoods, so that the $\alpha'$ expansion ceases to be valid. The resulting string effects are then represented effectively by a localized $\mathrm{OF1}$ source with lowest-derivative couplings. Conversely, upon resolving the orbifold, the explicit fixed-locus source disappears, while its charge and tension are encoded by the curvature of the smooth geometry through the $\mathcal{X}_8(R)$ and $\mathcal{E}_8(R)$ terms. It is in this sense that the $\mathrm{OF1}$ is geometrized.
A closely analogous phenomenon occurs for the D3-brane tadpole in F-theory. In the weakly coupled type IIB description, D7-branes and O7-planes are explicit localized sources whose higher-curvature worldvolume couplings induce D3-brane charge. In the dual M/F-theory description, the seven-brane system is geometrized by the elliptic fibration, while its total induced D3-brane charge is encoded by the bulk coupling $C_3\wedge \mathcal{X}_8(R)$ on the Calabi--Yau fourfold $Y_4$, yielding the topological contribution $\chi(Y_4)/24$ to the D3-brane tadpole \cite{Sethi:1996es}. Thus, in both cases, a charge represented by localized sources at a singular or weak-coupling locus is carried by higher-curvature terms in the corresponding smooth geometric description.

\subsection{Effective 2D $\mathcal{N}=(1,1)$ supergravity and vacua}
\label{sec:2DSUGRAex}

A torsion-free compact $\mathrm{Spin}(7)$ space admits one covariantly constant real chiral spinor and gives two real supercharges in type IIA compactifications to 2D \cite{JoyceBook2000,Becker:2000jc}. More precisely, since the two 10D supersymmetry parameters of type IIA have opposite chirality, the resulting 2D supercharges have opposite chiralities. Thus our explicit model preserves minimal non-chiral supersymmetry, namely 2D $\mathcal{N}=(1,1)$.  Crucially, the $\mathrm{OF1}$ quotient does not impose a further independent projection and therefore does not halve this supersymmetry.\footnote{Introducing $\mathrm{ON5}$ will break this to the actual minimal theory, namely $\mathcal{N}=(1,0)$.} The general structure of this theory is reviewed in appendix \ref{app:2DSUGRAgen}. Here, we specialize it to the setup of interest.

The action of 2D $\mathcal{N}=(1,1)$ supergravity is given by
\begin{equation}
S = 2\pi \int d^2x \sqrt{-g_2}\left(\frac{1}{2} J(q) R_2 +2 \mathcal{K}_{IJ}\partial^\mu q^I \partial_\mu q^J -V +ferm.\right),
\end{equation}
where the scalar potential is 
\begin{equation}
\label{eq:V2DSUGRA}
V = -\frac{1}{8} \mathcal{K}^{IJ}W_I W_J + \frac{1}{8}   \frac{(4W-J_I \mathcal{K}^{IL}W_L)^2}{J_I \mathcal{K}^{IK}J_K}.
\end{equation}
Here, $J(q)$, $\mathcal{K}_{IJ}(q)$ and $W(q)$ are arbitrary real functions of the scalar fields and $W_I = \frac{\partial}{\partial q^I}W$, $J_I = \frac{\partial}{\partial q^I}J$.
With respect to the appendix we added an overall $2 \pi$ factor to be in agreement with the units used in the microscopic setup. 
To match with our action \eqref{eq:S2Ddimred} arising from dimensional reduction, we first send
\begin{align}
g_{2, \mu\nu} \to \tilde g_{2,\mu\nu}, \qquad R_2 \to \tilde R_2,
\end{align}
in order to avoid rescaling $V$ by $L^2$, and then choose
\begin{align}
J(q) &= 2\rho^4 = 2 e^{2(y-\phi)},\\
\mathcal{K}_{IJ} &=-\frac{\rho^4}{4}\mathcal{G}_{IJ}.
\end{align}
Hence, one can calculate that $J_I \mathcal{K}^{IL}J_L = \frac{64}{7}\rho^4 \neq0$, as required by the formula \eqref{eq:V2DSUGRA} (the origin of this condition is explained in the appendix \ref{app:2DSUGRAgen}: it is needed to solve the equation of motion of the auxiliary field of the gravity multiplet), and also $J_I \mathcal{K}^{IJ}W_J=\frac{16}{7}W_y$. The non-trivial bit is to find the correct real superpotential, $W(q)$, that reproduces our desired scalar potential \eqref{eq:V_flux_source_y_bulk}, once inserted into the supergravity expression \eqref{eq:V2DSUGRA}, which is now
\begin{equation}
V = \frac{1}{2\rho^4}\mathcal{G}^{IJ}W_I W_J + \frac{7}{32\rho^4}\left(W-\frac{4}{7}W_y\right)^2.
\end{equation}
An educated guess can be formulated on the basis of our schematic formula \eqref{eq:Wcal2D}. However, one can readily check that including only $F_4$ flux is not enough to reproduce \eqref{eq:V_flux_source_y_bulk}, which contains contributions also from $F_0$ and $F_8$, see \eqref{eq:F8_norm_bulk_section} and \eqref{eq:F0_norm_bulk_section} respectively. Fortunately, the resolution is simple: one can notice that, besides $\Phi_4$, there are also the trivial calibrations $1$ (0-form scalar) and ${\rm dvol}_8$ (8-form volume). One can thus generalize the schematic formula \eqref{eq:Wcal2D} to
\begin{equation}
W \simeq  \int_8 F_{\rm even} \wedge \mathcal{P}(y) , \qquad \mathcal{P}(y) = e^{-y} + \Phi_4 + e^y {\rm dvol}_8,
\end{equation}
where $\mathcal{P}(y)$ is a polyform generalizing the non-trivial Cayley calibration $\Phi_4$ and $F_{\rm even}$ is a formal combination of all even-degree fluxes. 
Performing the integral, we get
\begin{equation}
W= L\rho^2 e^{\phi/4}\left[ \sum_{A=1}^7\left(N_Ae^{-X_A}+M_Ae^{X_A}\right)+\left(m e^y+n_8 e^{-y}\right)\right],
\end{equation}
which is the desired superpotential for our toroidal orbifold. Notice that it can be written in the manifestly geometric form
\begin{equation}
W=L\rho^2 e^{\phi/4}\left[\int_{8}F_4\wedge\Phi_4+m\,e^y+n_8\,e^{-y}\right],
\label{eq:W_geometric}
\end{equation}
which makes no reference to the toroidal structure; we exploit this in section~\ref{sec:generic}.
 Indeed, one can check that, also thanks to the overall normalization inserted to this purpose, this $W$ reproduces \eqref{eq:V_flux_source_y_bulk} via \eqref{eq:V2DSUGRA}. Crucially, for the matching to occur, one has to use the tadpole constraint \eqref{eq:tadpole_six_bulk}, since from \eqref{eq:V2DSUGRA} one really gets
\begin{equation}
V =\frac{1}{2} e^{\phi/2}L^2 \left[\sum_{A=1}^7\left( M_A e^{X_A}-N_A e^{-X_A} \right)^2+\left(m e^y - n_8 e^{-y}\right)^2 \right],
\end{equation}
and the tadpole \eqref{eq:tadpole_six_bulk} trades all the mixed terms in the squares above for $T_{\mathrm{OF1}}$.

We now investigate the vacuum structure of the 2D effective supergravity. We restrict our attention to supersymmetric vacua that are either anti-de Sitter (AdS) or Minkowski. As explained in the appendix \ref{app:2DSUGRAgen}, supersymmetric AdS vacua are determined by 
\begin{equation}
W=0, \qquad W_I = -J_I \mathcal{A},
\label{eq:FtermSUSYADS}
\end{equation}
where the explicit expression for $\mathcal{A}$ is given in \eqref{eq:appauxAEOM}. Crucially, $\mathcal{A}$ also determines the Ricci scalar via the integrability condition \eqref{eq:integrabilitySUGRA2D}, namely
\begin{equation}
R_2 = -2 \mathcal{A}^2.
\end{equation}
In particular, AdS requires $\mathcal{A}\neq 0$.
Supersymmetric Minkowski vacua are instead solutions of the equations 
\begin{equation}
W=0, \qquad W_I = 0,
\end{equation}
and indeed they feature $\mathcal{A}=0$.
To solve these supersymmetric conditions, it is useful to separate the non-vanishing overall prefactor from the superpotential and write
\begin{equation}
 W=\mathcal Y(y,\phi)\,\mathcal W(X_A,y),
 \qquad
 \mathcal Y=L\rho^2e^{\phi/4}=L e^{y-3\phi/4},
\label{eq:W_prefactor_bulk}
\end{equation}
where
\begin{equation}
 \mathcal W
 =\sum_{A=1}^7\left(N_Ae^{-X_A}+M_Ae^{X_A}\right)
 +m e^y+n_8e^{-y}.
\label{eq:reduced_W_bulk}
\end{equation}
The derivatives entering the supersymmetry conditions are therefore
\begin{align}
 W_{X_A}
 &=\mathcal Y\left(M_Ae^{X_A}-N_Ae^{-X_A}\right),
\label{eq:WX_susy_bulk}\\
 W_\phi&=-\frac{3}{4}W,
\label{eq:Wphi_susy_bulk}\\
 W_y
 &=W+\mathcal Y\left(m e^y-n_8e^{-y}\right),
\label{eq:Wy_susy_bulk}
\end{align}
and we have to ask $\mathcal Y\neq0$ for a finite-volume, finite-coupling vacuum, implying that $W=0$ becomes
\begin{equation}
\mathcal{W}=0.
\end{equation}

Imposing $W=0$ sets immediately $W_\phi=0$. In turn, inserting this into the second condition in \eqref{eq:FtermSUSYADS} tells us that $J_\phi \mathcal{A}=0$. Since $J_\phi=-4 \rho^4 = -4 \mathcal{Y}^2 e^{-\phi/2}/L^2\neq 0$, this forces $\mathcal{A}=0$. Hence, we have shown that our model does not have supersymmetric AdS vacua. We thus focus on Minkowski vacua.

 Imposing $W_{X_A}=0$, together with $\mathcal Y\neq 0$, fixes the shape moduli in terms of the flux integers
 \begin{equation}
 \label{eq:shape_susy_solution_bulk}
 e^{2X_A} = \frac{N_A}{M_A},
 \end{equation}
implying that $N_A M_A>0$.
Imposing $W=0$ leads automatically to $W_\phi=0$ and, combined with $W_y=0$, fixes the scalar $y$ in terms of the flux integers
\begin{equation}
\label{eq:y_susy_solution_bulk}
e^{2y} =\rho^4 e^{2\phi}= \frac{n_8}{m},
\end{equation}
implying that $m\, n_8>0$. Control over the supergravity approximation means the string frame volume should be parametrically large and hence $\frac{n_8}{m} \gg 1$. Yet the ratio of these fluxes will be fundamentally bound by the tadpole set by $\chi/24$. We quantify this bound, which depends on the total tadpole only, in section~\ref{sec:generic}. 

We still have to impose the tadpole condition  \eqref{eq:tadpole_six_bulk}. Let us set
\begin{equation}
c_0 = mn_8, \quad c_A = M_A N_A, \quad \epsilon_0 = {\rm sgn}(m) =  {\rm sgn}(n_8), \quad \epsilon_A = {\rm sgn}(M_A) = {\rm sgn} (N_A). 
\end{equation}
Introducing the collective index $i=(0,A)$, we can rewrite the tadpole  \eqref{eq:tadpole_six_bulk} and the condition $\mathcal{W}=0$ as
\begin{equation}
\sum_{i=0}^7 c_i = 14, \qquad \sum_{i=0}^7 \epsilon_i \sqrt{c_i}=0, \qquad \, c_i > 0.
\label{eq:tadci}
\end{equation}

At this stage the $c_i$ are positive real numbers and \eqref{eq:shape_susy_solution_bulk}--\eqref{eq:tadci} describe candidate vacua before flux quantisation is imposed. As we show below, the conditions \eqref{eq:corrected_quantisation} exclude every purely untwisted candidate in which all seven shape moduli are stabilised. We nevertheless retain these relations because they exhibit the structure of the would-be vacua and allow us to extract the generic features discussed in section~\ref{sec:generic}.

In any such candidate vacuum, all seven shape fields and the universal combination $y$ are fixed and the quadratic potential is positive in all propagating directions. In particular,
 \begin{equation}
 \left(m_X^2\right)^A{}_B
 =\delta^A{}_B\frac{4e^{\phi/2}}{\rho^4}N_AM_A>0,
 \qquad
 m_{y}^2
 =\frac{4e^{\phi/2}}{\rho^4}mn_8>0,
\label{eq:physical_bulk_masses_susy}
\end{equation}
so all propagating scalar fields retained in the untwisted diagonal truncation would have positive mass squared. These masses are quoted with respect to the physical external metric $g_2=L^2\tilde g_2$; passing from $\tilde g_2$ to $g_2$ absorbs the overall factor $L^2$ of the potential \eqref{eq:V_flux_source_y_bulk}.

\subsubsection{Flux quantisation versus the tadpole}
\label{subsec:obstruction}

Supersymmetric stabilisation of all seven shape moduli requires $c_A=N_AM_A>0$ for every $A$, as follows from \eqref{eq:shape_susy_solution_bulk}. Combining this with \eqref{eq:corrected_quantisation} gives
\begin{equation}
c_{1,2,3}\ge 1\,,\qquad c_{5,6,7}\ge 4\,,\qquad c_4\ge 8\,,\qquad c_0=mn_8\ge 1\,,
\end{equation}
and hence
\begin{equation}
mn_8+\sum_{A=1}^7 N_AM_A\;\ge\;1+1+1+1+8+4+4+4\;=\;24\;>\;14=\frac{\chi(X_8)}{24}\,.
\label{eq:obstruction_inequality}
\end{equation}
Under the Page quantisation and cycle-lift assumptions stated previously, no purely untwisted supersymmetric Minkowski vacuum of this model can stabilise all untwisted moduli: the minimal flux cost implied by these assumptions exceeds the available tadpole by ten units.  Similar observations have been made in related contexts \cite{Collinucci:2008pf}, and more recently formalized  as the Tadpole Conjecture \cite{Bena:2020xrh}. After some partial checks we suspect that this incompatibility may persist across the finite family of affine $\mathbb Z_2^4$ quotients with the same untwisted cohomology. 

It is worth commenting on the fate of scalars that are not stabilised at tree level, as is the case here. With only two supercharges there are no non-renormalisation theorems protecting them, so quantum corrections are expected to generate a potential; the vacuum construction then faces the usual Dine--Seiberg problem \cite{Dine:1985he}, since the corrections arise in an expansion whose control parameters are themselves among the unstabilised fields. In the unlikely event that some non-compact scalars remain exact moduli, two-dimensional infrared physics poses a further problem: the Green function of a free massless boson grows logarithmically at large separation, so field differences fluctuate without bound in the deep infrared and no conventional vacuum with a sharply defined expectation value exists \cite{Coleman1973}. One possible interpretation is that these large fluctuations drive the system towards decompactification, away from the 2D vacuum \cite{Banks:1995ci}. Massive fields are not affected by this issue, their correlators being cut off at separations of order the inverse mass, which is one more reason to insist on full classical stabilisation.

\subsubsection{No large volume regime}
\label{subsubsec:nolargevolume}
It is evident that, after flux quantization, no such vacuum as those considered above can achieve a parametrically large internal volume in string units, so $\alpha'$ corrections cannot be parametrically suppressed. Even for fluxless compactifications on compact $\mathrm{Spin}(7)$ manifolds, the internal metric receives $\alpha'$ corrections, although supersymmetry can be preserved by correcting the metric order by order \cite{Becker:2014rea}. For our flux compactifications, the conditions \eqref{eq:shape_susy_solution_bulk} and \eqref{eq:y_susy_solution_bulk} tell us that the tadpole \eqref{eq:tadpole_six_bulk} is a sum of positive terms adding up to fourteen, as in \eqref{eq:tadci}. To engineer large volume we should be able to implement a hierarchy such as $|N_A| \gg |M_A|$ and $|n_8| \gg |m|$, but this is prohibited by the tadpole.

One may wonder whether this conclusion can be avoided by relaxing supersymmetry and allowing cancellations in \eqref{eq:tadpole_six_bulk}. However, this does not help for Minkowski vacua in the present setup. The scalar potential depends on the fluxes through positive quadratic combinations. Extremizing with respect to the shape moduli and to $y$ gives, independently of supersymmetry,
\begin{equation}
e^{2X_A}=\left|\frac{N_A}{M_A}\right|,\qquad
e^{2y}=\left|\frac{n_8}{m}\right| .
\end{equation}
At such a critical point the flux energy is therefore proportional to $|m n_8|+\sum_{A=1}^7 |N_A M_A|$, but the Minkowski condition then requires $|m n_8|+\sum_{A=1}^7 |N_A M_A|=14$.
Together with the tadpole \eqref{eq:tadpole_six_bulk}, this forbids cancellations among large fluxes. Thus, even away from the supersymmetric locus, the equations of motion prevent a parametric large-volume regime within this simple $F_0$, $F_4$, $F_8$ flux and $\mathrm{OF1}$ setup.

\section{The solution from 10D}
\label{sec:10Dsolution}

To complete our analysis, we derive directly from the 10D type IIA Killing spinor equations the supersymmetry conditions used in section~\ref{sec:2DSUGRAex}.  We work in the same setup as in section~\ref{sec:toroidal_bulk_moduli}: the external spacetime is maximally symmetric, the internal metric is given in \eqref{eq:diag_metric_bulk} (supplemented with the rescaling $dx^i \to \sqrt 2 dx^i$ motivated in the previous section), the dilaton is constant, $H_3=0$, and the only non-vanishing RR fluxes are $F_0$, $F_4$ and $F_8$.  We also take the warp factor to be constant.  The calculation is therefore a bulk calculation in the unwarped smeared approximation.  A localised solution would in general require a non-trivial warp factor and dilaton profile.

It is useful to initiate the calculation in string frame, where the democratic type IIA supersymmetry variations have a particularly compact form, and translate the result at the end to the Einstein-frame variables $(X_A,y,\phi)$ used in section~\ref{sec:2DSUGRAex}.

\subsection{Gamma matrices, spinors and form conventions}

We use a mostly-plus 10D metric.  Curved 10D indices are denoted by $M,N=0,\ldots,9$, external 2D indices by $\mu,\nu=0,1$, and internal 8D indices by $m,n=1,\ldots,8$.  The gamma matrices obey
\begin{equation}
 \{\Gamma_M,\Gamma_N\}=2g_{MN},
 \qquad
 \{\check{\gamma}_\mu,\check{\gamma}_\nu\}=2g_{2,\mu\nu},
 \qquad
 \{\gamma_m,\gamma_n\}=2g_{8,mn},
\end{equation}
and antisymmetrised products have unit weight, for example
$\gamma_{mn}=\gamma_{[m}\gamma_{n]}$.  We choose the following decomposition compatible with the Clifford algebra,
\begin{equation}
 \Gamma_\mu=\check{\gamma}_\mu\otimes \mathbf 1,
 \qquad
 \Gamma_m=\check{\gamma}_*\otimes\gamma_m,
 \qquad
 \Gamma_{11}=\check{\gamma}_*\otimes\gamma_9,
 \label{eq:gamma_decomp_appendix}
\end{equation}
where $\check{\gamma}_*=\check{\gamma}^0\check{\gamma}^1, \check{\gamma}_*^2=\gamma_9^2=\mathbf 1$ and
$\gamma_9=\gamma_1\cdots\gamma_8$.  An internal $p$-form acts on an 8D spinor according to
\begin{equation}
 G_p\cdot\chi  =  \frac{1}{p!}G_{m_1\ldots m_p} \gamma^{m_1\ldots m_p}\chi = \slashed{G}_p \chi,
 \label{eq:clifford_action_appendix}
\end{equation}
and we denote by $\sigma$ the reversal operator on forms,
\begin{equation}
 \sigma(G_p)=(-1)^{\frac{1}{2}p(p-1)}G_p.
 \label{eq:sigma_operator_appendix}
\end{equation}

A torsion-free $\mathrm{Spin}(7)$ structure implies that the Riemannian holonomy is contained in $\mathrm{Spin}(7)$ and therefore guarantees at least one covariantly constant real chiral spinor.
If the Riemannian holonomy ${\rm Hol}(g)$ is exactly $\mathrm{Spin}(7)$, rather than a proper subgroup thereof, the space of parallel real spinors is one-dimensional, so that $\eta$ is unique up to normalization \cite{WangParallelSpinors,Becker:2000jc}. Theorem C of \cite{Joyce1996} states that, for a compact, simply-connected torsion-free Spin(7) manifold $X_8$, ${\rm Hol}(g)= \mathrm{Spin}(7)$ if the $\hat A$-genus is $\hat A(X_8)=1$. Since in terms of the Betti numbers we have $24 \hat A(X_8)=-1+b_1-b_2+b_3+b^+_4-2b^-_4$, we see that for the smooth resolved manifold studied in section \ref{sec:smoothres}, $\hat A(X_8)=1$. We thus have precisely one covariantly constant real chiral spinor $\eta$, unique up to normalization.
We choose its norm and chirality as
\begin{equation}
 \eta^{\mathrm T}\eta=1,
 \qquad
 \nabla_m\eta=0,
 \qquad
 \gamma_9\eta=\eta.
 \label{eq:eta_properties_appendix}
\end{equation}
The Cayley form is defined by the spinor bilinear \cite{Becker:2000jc}
\begin{equation}
 \Phi_{mnpq}=\eta^{\mathrm T}\gamma_{mnpq}\eta.
 \label{eq:cayley_bilinear_appendix}
\end{equation}
With the orientation for which $\star_8\Phi_4=\Phi_4$, this convention gives
\begin{equation}
 \Phi_4\cdot\eta=14\eta,
 \qquad
 \mathrm{dvol}_8\cdot\eta=\eta,
 \qquad
 \mathrm{dvol}_8\cdot\gamma_m\eta=-\gamma_m\eta.
 \label{eq:cayley_clifford_appendix}
\end{equation}
Indeed, we have $\Phi_4 \cdot \eta = \alpha \eta$, for some $\alpha$ to be determined. Acting with $\eta^T$ and using that $\eta$ has unit norm gives $\alpha=\eta^T(\Phi_4 \cdot \eta) = \frac{1}{4!}\Phi_{mnpq}\Phi^{mnpq}=14$. Similarly, from ${\rm dvol}_8 \cdot \eta = \frac{1}{8!}\epsilon_{m_1\dots m_8}\gamma^{m_1 \dots m_8} \eta$, we see that ${\rm dvol}_8 \cdot \eta = \gamma_9 \eta = \eta$ and ${\rm dvol}_8 \cdot \gamma_m \eta = \gamma_9 \gamma_m\eta =-\gamma_m \gamma_9\eta = -\gamma_m\eta$.

The type IIA supersymmetry parameters are Majorana--Weyl spinors of opposite 10D chirality. 
The $\mathrm{Spin}(7)$ metric    leaves one positive-chirality internal singlet. The $\mathrm{OF1}$ inversion acts on this spinor as $\gamma_9$ and therefore leaves it invariant, rather than imposing a further independent projection. Consequently, the $\mathrm{OF1}$ does not reduce the supersymmetry beyond that of the compactification: the combined background preserves $\mathcal{N}=(1,1)$ supersymmetry in two dimensions.
In accordance with the spinor conventions of appendix~\ref{app:2DSUGRAgen}, we package the two opposite-chirality supersymmetry parameters into a single 2D Majorana spinor,  $\zeta= P_+\zeta + P_-\zeta$, with $P_\pm=\frac{1}{2}(1\pm\check{\gamma}_*)$. The two 10D supersymmetry parameters are then
\begin{equation}
\epsilon_1=P_+\zeta\otimes\eta,
\qquad
\epsilon_2=P_-\zeta\otimes\eta.
\label{eq:iia_spinor_decomp_appendix}
\end{equation} 
Thus, the two external Weyl spinors are the two chiral projections of one 2D Majorana spinor and each $P_\pm\zeta$ has one real component.  This means that the Ansatz preserves two real supercharges of opposite 2D chirality.  The relative sign between $\epsilon_1$ and $\epsilon_2$ has been fixed so that the RR conditions below match the flux signs used in section~\ref{sec:2DSUGRAex}; reversing this relative sign reverses all RR fluxes simultaneously and does not change the classification of vacua.\footnote{That reversing the relative sign between $\epsilon_{1,2}$ can be compensated by reversing the sign of the RR fluxes can be seen from the supersymmetry transformations of the fermions, given below.}

In string frame and democratic formulation \cite{Bergshoeff:2001pv}, the supersymmetry variations of type IIA supergravity can be written as (we follow the conventions of \cite{Koerber:2010bx})
\begin{align}
 \delta\lambda^1&=
 \left(\slashed\partial\phi+\frac{1}{2}\slashed H\right)\epsilon_1
 -\frac{1}{16}e^\phi\Gamma^M\slashed F^{\rm tot}\Gamma_M\epsilon_2,
 \label{eq:iia_dilatino1_appendix}\\
 \delta\lambda^2&=
 \left(\slashed\partial\phi-\frac{1}{2}\slashed H\right)\epsilon_2
 -\frac{1}{16}e^\phi\Gamma^M\sigma(\slashed F^{\rm tot})\Gamma_M\epsilon_1,
 \label{eq:iia_dilatino2_appendix}\\
 \delta\psi^{1}_M&=
 \left(\nabla_M+\frac{1}{4}H_M\right)\epsilon_1
 -\frac{1}{16}e^\phi\slashed F^{\rm tot}\Gamma_M\epsilon_2,
 \label{eq:iia_gravitino1_appendix}\\
 \delta\psi^{2}_M&=
 \left(\nabla_M-\frac{1}{4}H_M\right)\epsilon_2
 -\frac{1}{16}e^\phi\sigma(\slashed F^{\rm tot})\Gamma_M\epsilon_1,
 \label{eq:iia_gravitino2_appendix}
\end{align}
where
\begin{equation}
 \slashed H=\frac{1}{3!}H_{MNP}\Gamma^{MNP},\qquad
 H_M=\frac{1}{2}H_{MNP}\Gamma^{NP},\qquad
 \slashed F^{\rm tot}=\sum_p\frac{1}{p!}F^{\rm tot}_{M_1\ldots M_p}\Gamma^{M_1\ldots M_p},
\end{equation}
and $F^{\rm tot}$ is the democratic RR polyform.  

Notice that, prior to introducing the supersymmetry transformations, our analysis did not assume any particular (Einstein or string) frame for the spacetime metric. To compare with section \ref{sec:universal_constraints}, it is convenient, but not necessary, to pretend that we have tacitly assumed to be working with the Einstein frame unit volume metric introduced in section~\ref{sec:universal_constraints}. However, now we have introduced Lagrangian information, in the form of the supersymmetry transformation, and we have switched to string frame, to work out the needed Killing spinor equations. To compare with section~\ref{sec:universal_constraints} and \ref{sec:toroidal_bulk_moduli}, we will have to go back to the previous unit-volume Einstein frame. We will do this further below, but for the time being we remain in string frame. 

2D maximal symmetry allows both purely internal magnetic components and electric components proportional to the external volume form. We therefore write
\begin{equation}
 F^{\rm tot}=F+\mathrm{dvol}_2\wedge F^{\rm el},
 \qquad
 F=F_0+F_4+F_8,
 \label{eq:democratic_flux_split_appendix}
\end{equation}
with $F^{\rm el}$ fixed by the 10D democratic duality constraint
\begin{equation}
\label{eq:appdemconstrdual}
 F^{\rm el} = \star_8^{(s)} \sigma(F),
 \end{equation}
  where we denoted with $(s)$ the Hodge operator with respect to the string frame metric. Besides, we have $F^{\rm tot} = \star_{10}^{(s)}\sigma(F^{\rm tot})$ or equivalently $F_p^{\rm tot} = (-1)^{\frac{(p-1)(p-2)}{2} }\star_{10}^{(s)} F_{10-p}^{\rm tot}$.
We choose the 2D orientation such that $ \slashed{\mathrm{dvol}}_2=\check\gamma_*$. From the identity $\bigl(\star^{(s)}_8\sigma(F_p)\bigr)\cdot\eta
 =F_p\cdot(\gamma_9\eta)$, combined with the democratic constraint \eqref{eq:appdemconstrdual}, we also find
\begin{equation}
 \slashed F^{\rm el}\eta=\slashed F\eta,
 \qquad
 \slashed F^{\rm el}\gamma_m\eta
 =-\slashed F\gamma_m\eta.
 \label{eq:Fel_actions_appendix}
\end{equation}
Since magnetic fluxes in the present truncation have degree $0$, $4$ or $8$,
$\sigma(F)=F$.  On the other hand, the electric partners have degrees $2$, $6$ and $10$, and therefore acquire a minus sign under $\sigma$.  It follows that
\begin{align}
 \slashed F^{\rm tot}  =\mathbf 1\otimes\slashed F   +\check{\gamma}_*\otimes\slashed F^{\rm el}, \qquad
 \sigma(\slashed F^{\rm tot}) =\mathbf 1\otimes\slashed F    -\check{\gamma}_*\otimes\slashed F^{\rm el}.
 \label{eq:sigmaFtot_clifford_split_appendix}
\end{align}

\subsection{Reduction of the 10D Killing spinor equations}

We now reduce the 10D supersymmetry equations to 2D. 
For a maximally symmetric 2D background, the Killing-spinor
equation can be written compactly as
\begin{equation}
\nabla_\mu P_\pm \zeta =\kappa_s\,\check{\gamma}_\mu P_\mp\zeta,
\label{eq:2d_killing_majorana_appendix}
\end{equation}
for a certain $\kappa_s$ which is constant on the background. 
The integrability condition follows from
\begin{equation}
\begin{aligned}
\bigl[\nabla_\mu,\nabla_\nu\bigr]\zeta &= \kappa_s^2 \left(\check{\gamma}_\nu\check{\gamma}_\mu-\check{\gamma}_\mu\check{\gamma}_\nu\right)\zeta=-2\kappa_s^2\check{\gamma}_{\mu\nu}\zeta =\frac{1}{4}R_{\mu\nu\rho\sigma} \check{\gamma}^{\rho\sigma}\zeta.
\end{aligned}
\end{equation}
Using that for a maximally symmetric 2D spacetime $R_{\mu\nu\rho\sigma} =\frac{1}{2}R_2 \left( g_{\mu\rho}g_{\nu\sigma} -g_{\mu\sigma} g_{\nu\rho} \right)$, implying in turn $\frac{1}{4} R_{\mu\nu\rho\sigma}\check{\gamma}^{\rho\sigma} = \frac{1}{4} R_2 \check{\gamma}_{\mu\nu}$, we find
\begin{equation}
R_2=-8\kappa_s^2.
\label{eq:2d_killing_curvature_appendix}
\end{equation}
Recall that $R_2$ is here still in string frame. 

We next insert \eqref{eq:iia_spinor_decomp_appendix} into the gravitino
variations.  The identities from the previous section give
\begin{align}
 \slashed F^{\rm tot}\Gamma_\mu\epsilon_2 &=2\check\gamma_\mu P_-\zeta\otimes\slashed F\eta,  &  \sigma(\slashed F^{\rm tot})\Gamma_\mu\epsilon_1 &=2\check\gamma_\mu P_+\zeta \otimes\slashed F\eta,
 \label{eq:external_RR_actions_appendix}\\
 \slashed F^{\rm tot}\Gamma_m\epsilon_2  &=-2 P_-\zeta\otimes\slashed F\gamma_m\eta, & \sigma(\slashed F^{\rm tot})\Gamma_m\epsilon_1  &=2 P_+\zeta \otimes\slashed F\gamma_m\eta.
 \label{eq:internal_RR_actions_appendix}
\end{align}
Consequently, the two external gravitino equations, with also $H_3=0$, coincide and reduce to
\begin{equation}
 \delta \psi^{1,2}_\mu=0 \qquad \Rightarrow \qquad \kappa_s\,\eta-\frac{1}{8}e^\phi\slashed F\eta=0.
 \label{eq:external_gravitino_reduced_appendix}
\end{equation}
This equation exhibits explicitly the RR gravitino shift.  
 The two internal gravitino equations also coincide and give
\begin{equation}
 \delta \psi^{1,2}_m=0 \qquad \Rightarrow \qquad \slashed F\gamma_m\eta=0.
 \label{eq:internal_gravitino_reduced_appendix}
\end{equation}

The dilatino equation contains a different linear combination of the RR
fields.  Using the gamma matrix identity in general $D$ dimensions $\gamma_\alpha \gamma^{\beta_1\beta_2 \dots \beta_n}\gamma^\alpha = (-1)^n(D-2n)\gamma^{\beta_1 \beta_2 \dots \beta_n}$, we can calculate
\begin{align}
 \Gamma^M\slashed F^{\rm tot}\Gamma_M\epsilon_2
 &=4P_-\zeta\otimes
 \left(5\slashed F_0+\slashed F_4-3\slashed F_8\right)\eta,
 \label{eq:dilatino_action1_appendix}\\
 \Gamma^M\sigma(\slashed F^{\rm tot})\Gamma_M\epsilon_1
 &=4P_+\zeta\otimes
 \left(5\slashed F_0+\slashed F_4-3\slashed F_8\right)\eta,
 \label{eq:dilatino_action2_appendix}
\end{align}
and thus both dilatino equations for $H_3=0 = \partial \phi$ become
\begin{equation}
 \delta \lambda^{1,2}=0 \qquad \Rightarrow \qquad  \left(5\slashed F_0+\slashed F_4-3\slashed F_8\right)\eta=0.
 \label{eq:dilatino_reduced_appendix}
\end{equation}

To summarize,  the independent 10D algebraic equations equivalent to $\mathcal{N}=(1,1)$  supersymmetry in 2D are
\begin{align}
\label{eq:KSeq1}
 \kappa_s\eta- \frac{1}{8}e^\phi\slashed F\eta &=0,\\
  \label{eq:KSeq2}
 \slashed F\gamma_m\eta&=0,\\
 \left(5\slashed F_0+\slashed F_4-3\slashed F_8\right) \eta&=0,
 \qquad
 \label{eq:KSeq3}
\end{align}
with $\slashed F = \slashed F_0 + \slashed F_4 + \slashed F_8$ in our setup.
These have been derived starting from 10D string frame, while our analysis in section~\ref{sec:2DSUGRAex} has been performed in Einstein frame and with a unit volume 8D metric. Before inserting our specific flux configurations into \eqref{eq:KSeq1}-\eqref{eq:KSeq3} and solving them, we recast them in the appropriate frame. 

To pass from the 10D string frame metric to the 10D Einstein frame metric we have the usual rescaling $g^s_{MN} = e^{\phi/2}g^E_{MN}$. Then, to have a unit-volume metric in the internal space, we have to rescale the 8D part of the metric further by $\rho$. We thus have
\begin{equation}
g^s_{2,\mu\nu} = e^{\frac{\phi}{2}} g^E_{2,\mu\nu}\,, \qquad  g^s_{8,mn} = e^{\frac{\phi}{2}} \rho \,\tilde g_{8,mn} = e^{\frac{y}{2}}\, \tilde g_{8,mn}
\end{equation}
and gamma matrices are rescaled with the square roots of these factors, $\check{\gamma}_\mu^s = e^{\frac{\phi}{4}}\check{\gamma}_\mu^E$ and $\gamma_m^s = e^{\frac{\phi}{4}}  \rho^{\frac12}\, \tilde \gamma_m = e^{\frac y4}\, \tilde \gamma_m $,  where we recall that $\rho^4= e^{2y-2\phi}$. Then, we need to rescale
\begin{equation}
\slashed F_p^s =\left(e^{\frac{\phi}{4}}  \rho^{\frac12}\right)^{-p}\,\tilde{ \slashed F_p} = e^{-\frac{y}{4}p} \tilde{ \slashed F_p} ,
\end{equation}
and the 2D Killing-spinor equation becomes 
\begin{equation}
\nabla_\mu P_\pm \zeta = \kappa_E \check\gamma_\mu^E P_\mp \zeta, \qquad \text{with} \quad \kappa_E = e^{\frac{\phi}{4}} \kappa_s.
\end{equation}
This gives the integrability condition $R_2^E = -8 \kappa_E^2$, matching with \eqref{eq:integrabilitySUGRA2D} for $\kappa_E = \frac{1}{2} \mathcal{A}$.
Performing these rescalings on \eqref{eq:KSeq1}-\eqref{eq:KSeq3}, we find (we omit the tildes on $\slashed F_p$ and $\gamma_m$ from now on; everything is written in the appropriate Einstein frame with unit-volume 8D metric)
\begin{align}
\label{eq:KSeqE1}
e^{-\frac{\phi}{4}}\kappa_E \eta - \frac{1}{8} \rho^{-2} \left(e^y \slashed F_0 + \slashed F_4 + e^{-y}\slashed F_8\right)\eta&=0,\\
\label{eq:KSeqE2}
(e^y \slashed  F_0 + \slashed F_4 + e^{-y} \slashed F_8) \gamma_m \eta &= 0,\\
\label{eq:KSeqE3}
 \left(5e^y\slashed F_0+\slashed F_4-3e^{-y}\slashed F_8\right) \eta&=0.
\end{align}

We finally arrived at the desired Killing spinor equations in the appropriate frame.
We now solve them in the flux basis of section~\ref{sec:2DSUGRAex}. It is nevertheless important to recall that they are not enough to determine a consistent background, as they must be supplemented by the Bianchi identities. In our case, these are just the tadpole constraint \eqref{eq:tadpole_six_bulk}. 

\subsection{Solution and matching with 2D supergravity}

Let us recall the expression for our fluxes from section \ref{sec:toroidal_bulk_moduli} (the formulae below are thus in the unit-volume Einstein frame)
\begin{equation}
F_0 = m, \qquad F_4 = \sum_{A=1}^7 \left(N_A \omega_A + M_A \tilde \omega_A\right), \qquad F_8 = n_8 {\rm dvol}_8.
\end{equation}
It is useful to introduce the (anti)-self dual 4-forms, $\star_8\Psi^\pm_A = \pm \Psi^\pm_A$, defined as 
\begin{equation}
\Psi^\pm_A = e^{X_A}\omega_A \pm e^{-X_A}\tilde \omega_A, \qquad \text{with} \qquad \Phi_4 = \sum_{A=1}^7 \Psi^+_A.
\end{equation}
From $\omega_A \wedge \tilde \omega_B = \delta_{AB}{\rm dvol}_8$, one has $\Psi^+_A\wedge \Psi^+_B = 2 \delta_{AB}{\rm dvol}_8$ and thus $\Psi_A^+\wedge \Phi_4 = \Psi_A^+\wedge \star_8 \Phi_4 = 2 {\rm dvol}_8$.
We can thus decompose $F_4$ into its (anti)-self dual parts
\begin{equation}
 F_4 =\frac{1}{2}\sum_{A=1}^7 \left[ \left(N_Ae^{-X_A}+M_Ae^{X_A}\right)\Psi_A^+ +\left(N_Ae^{-X_A}-M_Ae^{X_A}\right)\Psi_A^- \right] = F_4^+ + F_4^-.
 \label{eq:F4_pm_decomposition_appendix}
\end{equation}
This basis has the advantage that
\begin{equation}
\label{eq:appPsiAprop}
\Psi^-_A \cdot \eta = 0, \qquad \Psi_A^+ \cdot \gamma_m \eta = 0, \qquad   \Psi^+_A\cdot \eta = 2 \eta,
\end{equation}
 where the first and the second properties follow from applying the 4-form identity $(\star_8 G_4) \cdot \chi = G_4 \cdot \gamma_9 \chi$ to the case in which $\chi$ has positive chirality; the last one can be proved with a bit more work.\footnote{Following $e.g.$\cite{Karigiannis_2008}, on a Spin(7) manifold the space of 4-forms decomposes into self-dual and anti self-dual subspaces, $\Lambda^4 = \Lambda^4_+\oplus\Lambda^4_-$, with $\Lambda^4_+=\Lambda^4_1 \oplus \Lambda^4_7 \oplus\Lambda^4_{27}$ and $\Lambda^4_- = \Lambda^4_{35}$. Each subscript in each summand in $\Lambda^4_\pm$ is an irreducible Spin(7) representation. In particular,  $\Lambda_1^4$ is the unique (up to overall rescaling) Cayley form $\Phi_4$. On the other hand, the positive chirality fundamental Spin(8) representation, $\mathbf{8}^+$, decomposes into Spin(7) representations as $\mathbf{8}^+ = \mathbf{1}\oplus \mathbf{7}$, see $e.g.$ \cite{Munoz:2016rgc}. Since a 4-form Clifford multiplication of an even degree-form with a spinor preserves chirality of the spinor, and since $\eta \in \mathbf{8}^+$, then also $\Psi_A^+ \cdot \eta \in \mathbf{8}^+$. Both $\Psi_A^+$ and $\eta$ are orbifold-invariant, but there is only a unique invariant, positive chirality spinor preserved by our orbifold choice and thus it must be that $\Psi_A^+\cdot \eta = \alpha_A \eta$ for some coefficient $\alpha_A$. Since $\eta$ has unit norm, we have $\alpha_A = \frac{\eta^T \Psi_A^+\cdot \eta}{\eta^T \eta} = \int\Psi_A^+ \wedge \star_8 \Phi_4 = 2$, where we used \eqref{eq:cayley_bilinear_appendix}.
 Alternatively, since the  map $\Lambda^4_7 \to\mathbf{7} \subset \mathbf{8}^+$ is injective (in fact it is an isomorphism), $\Psi_A^+$ cannot have a piece in $\Lambda^4_7$, therefore it is of the form $\Psi_A^+ = \beta_A \Phi_4+ \chi_{A}$, for some $\chi_A \in \Lambda^4_{27}$ and some coefficient $\beta_A$. Orthogonality gives $\beta_A = \frac{\int \Psi_A^+ \wedge \star_8 \Phi_4}{\int \Phi_4 \wedge \star_8 \Phi_4} = \frac{2}{14}=\frac{1}{7}$. The Clifford map $\Lambda_+^4 \to \mathbf{8}^+$ kills the piece $\Lambda^4_{27}$ and thus we have $\Psi_A^+ \cdot \eta = \beta_A \Phi_4 \cdot \eta = \frac{1}{7} (14 \eta) = 2 \eta$, where we used \eqref{eq:cayley_clifford_appendix}.} 
Finally, recall that ${\rm dvol}_8 \cdot \eta = \eta$ and ${\rm dvol}_8 \cdot \gamma_m \eta = - \gamma_m \eta$.

Inserting these RR fluxes into the equations \eqref{eq:KSeqE1}-\eqref{eq:KSeqE3}, and using the above properties \eqref{eq:appPsiAprop} we obtain the following conditions
\begin{align}
 \label{eq:external_gravitino_condition_appendix}
 \delta\psi_\mu=0: &\qquad  8\rho^2 e^{-\frac{\phi}{4}}\kappa_E =m e^y+n_8e^{-y}+\sum_{A=1}^7\left(N_Ae^{-X_A}+M_Ae^{X_A}\right)\\
  \delta\psi_m=0: &\qquad m e^y-n_8e^{-y}=0, \qquad N_Ae^{-X_A}-M_Ae^{X_A}=0,
 \label{eq:internal_gravitino_solution_appendix}\\
 \delta\lambda=0: &\qquad  5m e^y-3n_8e^{-y} +\sum_{A=1}^7\left(N_Ae^{-X_A}+M_Ae^{X_A}\right)=0.
 \label{eq:dilatino_scalar_appendix}
\end{align}
The internal gravitino equation \eqref{eq:KSeqE2} contains two pieces, $[(me^y-n_8 e^{-y}) \mathbf{1} + \slashed F_4^{-}]\gamma_m \eta=0$. The first ($F_0, F_8$) is in the singlet representation of Spin(7), $i.e.$ the $\mathbf{1}$, while the second ($F_4^-$) is in the symmetric traceless representation, $i.e.$ the $\mathbf{35}$. As such, they must vanish independently. This is the reason we get two different conditions. 

These equations have a direct 2D interpretation.  Writing the superpotential of section~\ref{sec:2DSUGRAex} as
\begin{equation}
 W=\mathcal Y\,\mathcal W,  
 \end{equation}
 with
 \begin{equation}
   \mathcal Y=L\rho^2e^{\phi/4}\neq0,  \qquad \mathcal W  =\sum_{A=1}^7\left(N_Ae^{-X_A}+M_Ae^{X_A}\right) +m e^y+n_8e^{-y},
 \label{eq:W_comparison_appendix}
\end{equation}
its relevant derivatives are
\begin{equation}
 \frac{W_{X_A}}{\mathcal Y}
 =M_Ae^{X_A}-N_Ae^{-X_A},
 \qquad
 W_\phi=-\frac{3}{4}W,
 \qquad
 \frac{W_y}{\mathcal Y}
 =\mathcal W+m e^y-n_8e^{-y}.
 \label{eq:W_derivatives_comparison_appendix}
\end{equation}
We can thus build the following dictionary between 10D and 2D conditions
\begin{align}
 \delta\psi_\mu=0  &\quad\Longleftrightarrow\quad
 8\rho^2e^{-\frac{\phi}{4}}\kappa_E=\frac{W}{\mathcal Y}\\
\left.\delta\psi_m\right|_{\bf 1}=0  &\quad\Longleftrightarrow\quad W_y-W=0,\\
 \left.\delta\psi_m\right|_{\bf 35}=0  &\quad\Longleftrightarrow\quad W_{X_A}=0,\\
 \delta\lambda=0 &\quad\Longleftrightarrow\quad 4W_y-3W=0,
\end{align}
which implies also
\begin{equation}
\delta \lambda - 4 \delta \psi_m|_{\mathbf{1}} = 0 \quad \quad\Longleftrightarrow\quad W = 0 \quad \quad\Longleftrightarrow\quad W_\phi = 0.
\end{equation}
Therefore, from $\left.\delta\psi_m\right|_{\bf 1}=0$ one must also have $W_y=0$. 
Consequently, the external gravitino equation gives $\kappa_E = 0$ ($i.e.$ $\mathcal{A}=0$) and thus $R_2=0$, namely the 2D spacetime must be Minkowski. 
To summarize, we have shown that the 10D Killing spinor equations are thus equivalent to the following conditions on the function $W$,
\begin{equation}
W = 0, \qquad W_I=0, \qquad q^I=\{X_A, y, \phi\},
\end{equation}
which are precisely the conditions for supersymmetric Minkowski vacua of 2D $\mathcal{N}=(1,1)$ supergravity given in \eqref{eq:appSUSYMinkSUGRA}. 

For our toroidal model, these conditions combined with the tadpole \eqref{eq:tadpole_six_bulk} reproduce precisely the structure \eqref{eq:shape_susy_solution_bulk}--\eqref{eq:tadci} found in section~\ref{sec:2DSUGRAex}. The 10D Killing spinor equations therefore agree with the 2D $\mathcal N=(1,1)$ analysis of section~\ref{sec:2DSUGRAex}.

\section{\texorpdfstring{On general $\mathrm{Spin}(7)$ compactifications }{On general Spin(7) compactifications}}
\label{sec:generic}

The toroidal orbifold provides an explicit realisation of the metric and universal sectors, but these sectors admit a geometric formulation on any compact manifold $X_8$ with holonomy $\mathrm{Spin}(7)$. We consider the same $F_0$, $F_4$ and $F_8$ fluxes, with the tadpole cancelled by the curvature contribution $\chi/24$,
\begin{equation}
m\,n_8+\frac{1}{2}\int_{X_8}F_4\wedge F_4=\frac{\chi(X_8)}{24}\,.
\label{eq:tadpole_general}
\end{equation}
The universal sector, $J=2\rho^4$ and the $(y,\phi)$ kinetic block, is independent of the detailed internal geometry, while the superpotential takes the geometric form \eqref{eq:W_geometric}. Let $s^a$ parametrise the torsion-free $\mathrm{Spin}(7)$ structures at unit volume. The kinetic metric for these shape moduli is given by \cite{Becker:2003wb}
\begin{equation}
\mathcal G_{ab}=\frac{1}{2}\int_{X_8}\partial_a\Phi_4\wedge\star_8\,\partial_b\Phi_4\,,
\label{eq:L2metric}
\end{equation}
which reduces to $\delta_{AB}$ in the toroidal truncation.\footnote{According to the deformation theory of torsion-free $\mathrm{Spin}(7)$ structures \cite{Joyce1996}, tangent vectors to the unit-volume moduli space can be represented by harmonic anti-self-dual 4-forms $\partial_a\Phi_4$, which belong to the ${\bf 35}$ of $\mathrm{Spin}(7)$. The dimensional reduction of the Einstein--Hilbert term then expresses the kinetic metric in terms of the $\mathrm{Spin}(7)$-invariant quadratic form $\int_{X_8}\partial_a\Phi_4\wedge\star_8\partial_b\Phi_4$ \cite{Becker:2003wb}. Its normalization in our conventions can be fixed using the toroidal truncation. From \eqref{eq:normalized_cayley_bulk}, $\partial_{X_A}\Phi_4=e^{X_A}\omega_A-e^{-X_A}\widetilde\omega_A$, and \eqref{eq:hodge_star_pairs} together with \eqref{eq:quotient_unit_pairing} gives $\frac{1}{2}\int_{\mathbb T^8/\Gamma}\partial_{X_A}\Phi_4\wedge\star_8\partial_{X_B}\Phi_4=\delta_{AB}$. This reproduces the shape-moduli kinetic metric $\mathcal G_{AB}=\delta_{AB}$ appearing in \eqref{eq:S2Ddimred} and therefore fixes the normalization in \eqref{eq:L2metric}.
}
Since a fixed-volume variation of the Cayley form is represented by an anti-self-dual harmonic 4-form \cite{Joyce1996},
\begin{equation}
W_a=\mathcal Y\int_{X_8}F_4\wedge\partial_a\Phi_4=-\,\mathcal Y\int_{X_8}F_4^-\wedge\star_8\,\partial_a\Phi_4\,,
\end{equation}
and therefore $W_a=0$ for every shape direction is equivalent to $F_4^-=0$. Within the bulk approximation described above, we impose exactly the topological tadpole generated by $\mathcal{X}_8(R)$. We further make the assumption that the associated integrated contribution of $\mathcal{E}_8(R)$ is captured by the same BPS energy as in the orbifold description, fixed by $\chi(X_8)/24$ \cite{Gukov:2002iq}, while neglecting the remaining local higher-derivative corrections to the metric, effective action and supersymmetry transformations. Under this assumption, using the tadpole \eqref{eq:tadpole_general}, the potential for the metric and universal fields within our approximation can be written as
\begin{equation}
V=\frac{1}{2}L^2e^{\phi/2}\left[2\int_{X_8}F_4^-\wedge\star_8 F_4^-+\left(m\,e^y-n_8\,e^{-y}\right)^2\right].
\label{eq:V_geometric}
\end{equation}
The axionic sectors associated with $b_2,b_3\neq0$, which vanish in the untwisted truncation but are present on generic smooth $\mathrm{Spin}(7)$ manifolds, require a separate derivation.

The Killing spinor equations \eqref{eq:KSeqE1}--\eqref{eq:KSeqE3} were likewise derived without using the toroidal structure. A supersymmetric Minkowski vacuum in this sector, and under our assumptions, satisfies
\begin{equation}
F_4^-=0\,,\qquad m\,e^y=n_8\,e^{-y}\,,\qquad \int_{X_8}F_4\wedge\Phi_4=-2\epsilon_0\sqrt{mn_8}\,,
\label{eq:general_susy_conditions}
\end{equation}
where $\epsilon_0={\rm sgn}(m)={\rm sgn}(n_8)$. In particular,
\begin{equation}
e^{2y}=\rho^4e^{2\phi}=\frac{n_8}{m}\,.
\end{equation}
These equations imply a stronger universal bound than follows from positivity alone. Defining $\|G_4\|^2=\int_{X_8}G_4\wedge\star_8G_4$, the unit-volume Cayley form obeys $\|\Phi_4\|^2=14$. The last condition in \eqref{eq:general_susy_conditions} and the Cauchy--Schwarz inequality therefore give
\begin{equation}
4mn_8=\left|\int_{X_8}F_4\wedge\Phi_4\right|^2\leq \|F_4\|^2\|\Phi_4\|^2=14\|F_4\|^2\,,
\end{equation}
and hence
\begin{equation}
\|F_4\|^2\geq\frac{2mn_8}{7}\,.
\label{eq:F4_cauchy_bound}
\end{equation}
Since $F_4$ is self-dual at the Minkowski vacuum, the tadpole \eqref{eq:tadpole_general} becomes
\begin{equation}
\frac{\chi(X_8)}{24}=mn_8+\frac{1}{2}\|F_4\|^2\geq\frac{8}{7}mn_8\,.
\end{equation}
Finally, $m$ is a nonzero integer, so $e^{2y}=mn_8/m^2\leq mn_8$. Combining these results yields
\begin{equation}
e^{2y}\leq mn_8\leq\frac{7}{8}\frac{\chi(X_8)}{24}=\frac{7\chi(X_8)}{192}\,.
\label{eq:ybound}
\end{equation}
Thus parametrically suppressing $\alpha'$ corrections within this class requires a family of $\mathrm{Spin}(7)$ manifolds with parametrically large Euler characteristic.

The conditions \eqref{eq:general_susy_conditions} do not by themselves prove that all metric moduli are stabilised. For a fixed quantised class $[F_4]$, the self-duality locus can be empty or non-isolated, while the norm of the class is restricted by the tadpole. Consequently, counting the $b_4^-$ self-duality equations does not determine the flux cost of stabilising the $b_4^-$ shape moduli, and no general conclusion about an analogue of the tadpole problem \cite{Bena:2020xrh} follows from this counting argument.

Establishing the existence of candidate vacua on a smooth compact $\mathrm{Spin}(7)$ manifold therefore requires explicit control of its integral flux lattice, and of the locus where an allowed class becomes self-dual and satisfies the last condition in \eqref{eq:general_susy_conditions}. This remains an open geometric problem.

\section{Conclusions}
\label{sec:conclusions}

In this paper we initiated a study of tree-level type IIA flux compactifications to 2D on Ricci-flat spaces with $\mathrm{Spin}(7)$ holonomy. We first related the calibration forms of special-holonomy manifolds to supersymmetric brane and plane configurations on toroidal orbifolds. In 2D this leads naturally to the less familiar NS-charged defects $\mathrm{OF1}$ and $\mathrm{ON5}$ \cite{Witten:1998xy,Hanany:2000fq}. The $\mathrm{OF1}$ quotient used in our explicit construction is a spacetime orbifold and imposes no additional independent supersymmetry projection, so the $\mathcal N=(1,1)$ supersymmetry selected by the $\mathrm{Spin}(7)$ compactification survives.

We derived universal constraints on 2D flux vacua directly from the reduced 10D action. Since the Einstein--Hilbert term is topological in 2D, there is no ordinary Einstein frame, and the external length scale must be varied together with the internal volume and the dilaton. The resulting algebraic equations give the 2D analogue of the Maldacena--Nu\~nez obstruction \cite{Maldacena:2000mw,VanRiet:2023pnx}: in the absence of negative-tension sources, nontrivial internal flux supports only AdS$_2$ vacua, and under the assumptions of our analysis these vacua are not parametrically scale separated. Negative-tension sources are therefore required to evade the simplest no-go arguments.

We then constructed the 2D effective theory for a toroidal $\mathrm{Spin}(7)$ orbifold. The untwisted cohomology permits only $F_0$, $F_4$ and $F_8$ fluxes, whose fundamental string tadpole is cancelled by the $\mathrm{OF1}$ system with total charge $-\chi/24$ \cite{Sethi:1996es,Gukov:2002iq}. We derived the scalar kinetic terms and potential of the untwisted sector and embedded them into 2D $\mathcal N=(1,1)$ supergravity. The resulting real superpotential has a geometric expression in terms of the RR fluxes and the Cayley calibration, and its supersymmetry conditions agree with the 10D Killing spinor equations. Before imposing flux quantisation, the candidate Minkowski vacua fix all seven untwisted shape modes and the propagating volume--dilaton combination, while the orthogonal constant deformation is non-propagating.

Page flux quantisation changes this conclusion for the explicit model. A candidate stabilising all propagating untwisted scalars then requires at least $24$ tadpole units, whereas the resolved manifold supplies only $\chi/24=14$. Consequently, this model has no purely untwisted supersymmetric Minkowski vacuum that stabilises all untwisted modes. This obstruction applies to the purely untwisted flux sector; twisted flux components and more refined quantisation conditions \cite{Diaconescu:2000wy} remain to be analysed. It is possible that they may lift the obstruction.

For a general compact $\mathrm{Spin}(7)$ manifold, the metric and universal sectors retain our geometric form. At a supersymmetric Minkowski vacuum, the Cauchy--Schwarz inequality provides the volume bound $\mathcal{V}_s\leq 7\chi/192$, where $\mathcal{V}_s$ is the string-frame internal volume in units where \(2\pi\sqrt{\alpha'}=1\). Thus large Euler characteristic is necessary for parametrically suppressing $\alpha'$ corrections, although it is not sufficient \cite{Becker:2014rea}. An allowed quantised 4-form class must also become self-dual at an appropriate point in moduli space and satisfy a remaining singlet condition.

Next steps are to determine the full integral 4-form lattice of the resolved model, including exceptional cycles and a more precise flux quantisation, and to include twisted fluxes and the axionic sectors associated with nonzero $b_2$ and $b_3$. It would also be important to construct localised solutions with the corresponding warp factor and dilaton profile and to explore broader compactifications involving $\mathrm{ON5}$-planes, metric flux or intrinsic torsion. These extensions will determine whether controlled Minkowski or scale-separated AdS$_2$ vacua can arise in this framework.

\subsection*{Acknowledgments}
We would like to thank George Tringas for useful discussions and comments on an earlier draft of this paper. The work of N.C. is supported by the Research Foundation Flanders (FWO grant 1259125N). The work of A.H. is supported by Erasmus+ Grant (2025-1-TR01-KA131-HED-000308915). A.H. thanks the Department of Physics and Astronomy at the KU Leuven for kind hospitality during the completion of this work. The work of T.V.R. is supported by the KU Leuven C1 grant ZKE7799C16/25/010. The work of T.W. is supported in part by the NSF grant PHY-2210271. T.W. thanks the Aspen Center for Physics, which is supported by the National Science Foundation grant PHY-2210452, for hospitality during the final stages of this project.

\paragraph*{AI assistance.}
The authors used several large language models to assist with writing, editing, and checking calculations in this paper. The authors have carefully reviewed all such contributions and take full responsibility for the content, accuracy, and conclusions of the work.

\appendix

\section{\texorpdfstring{2D $\mathcal N=(1,1)$ supergravity}
{2D N=(1,1) supergravity}}
\label{app:2DSUGRAgen}

In this appendix, we recall the main properties of 2D $\mathcal{N}=(1,1)$ supergravity that we employ in section~\ref{sec:2DSUGRAex}.
We mainly follow \cite{Bergamin:2003am}, albeit with different conventions for the spacetime signature, which we take to be $(-+)$, and for the fermions, especially the Majorana condition.

We work with a 2D Clifford algebra $\{\check\gamma^a, \check\gamma^b\} = 2 \eta^{ab} \mathbf1_2$,  with $a=0,1$ flat spacetime indices, supplemented by the chirality matrix $\check\gamma_*= \check\gamma^0 \check\gamma^1$, such that $\check\gamma^2_*=\mathbf1_2$. We also define the chirality projectors $P_\pm = \frac{1}{2} \left(\mathbf1_2 \pm \check\gamma_*\right)$, implementing the Weyl condition on the fermions.
In 2D the minimal spinor is Majorana--Weyl and we combine the two chiral components into a Majorana spinor, $\psi^\alpha=(\psi^+,\psi^-)$. Spinor indices are raised and lowered with the matrix $C_{\alpha\beta}$,
\begin{equation}
 \psi_\alpha=C_{\alpha\beta}\psi^\beta,
 \qquad
 \psi^\alpha=C^{\alpha\beta}\psi_\beta, 
\label{eq:app-mp-index-conventions}
\end{equation}
such that $C_{\alpha\beta}=-C_{\beta\alpha}$, $C_{\alpha\beta} C^{\beta\gamma}=\delta_\alpha^\gamma$, and
giving $ \psi_+=-\psi^-$ and $\psi_-=\psi^+$. In our conventions, the Majorana condition is
\begin{equation}
\psi^c = B\psi^* = \psi,
\end{equation}
where $B$ is such that
\begin{equation}
 B^{-1}\check\gamma^aB=(\check\gamma^a)^*,  \qquad  B^*B=\mathbf1_2.
\end{equation}
We work with the real representation
\begin{equation}
\check\gamma^0=\begin{pmatrix}0&1\\ -1&0\end{pmatrix},
\qquad
\check\gamma^1=\begin{pmatrix}0&1\\ 1&0\end{pmatrix},
\qquad
\eta^{ab}=\operatorname{diag}(-1,+1),
\end{equation}
which in turn implies
\begin{equation}
B=i\mathbf 1_2, \qquad \psi^*=-i\psi, \qquad C=-\check\gamma^0.
\end{equation}
In this basis, the spinors are written in a phase-rotated Majorana frame: if $\psi_R$ has ordinary real components, $\psi^*_R = \psi_R$, we work with $\psi =e^{i\pi/4}\psi_R$, which is the reason why $\psi^* = -i \psi$ contains the additional factor $i$. We take complex conjugation to act as $(XY)^* = Y^* X^*$.
We also introduce the superspace coordinates $\Theta^\alpha=(\Theta^+,\Theta^-)$, such that  $\Theta^2  = \Theta^\alpha \Theta_\alpha = 2 \Theta^- \Theta^+$. Since $(\Theta^\alpha)^*=-i\Theta^\alpha$, the bilinear $\Theta^2$ is real. The superspace integration is performed via
\begin{equation}
{\rm d}^2 \Theta = {\rm d} \Theta^+ {\rm d}\Theta^-, \qquad \int {\rm d^2} \Theta \,\Theta^- \Theta^+ = 1, \qquad \frac{1}{2} \int {\rm d^2} \Theta  \,\Theta^2 = 1.
\end{equation}
Next, we introduce the superfields of interest for us.

The minimal off-shell supergravity multiplet is $(e_\mu{}^a,\psi_\mu,\mathcal{A})$, where $e_\mu{}^a$ is the zweibein, $\psi_\mu$ is a Majorana gravitino and $\mathcal{A}$ is a real auxiliary scalar. It is encoded in the superspace Berezinian $E$ and in a real curvature superfield $\mathcal R$. Their superspace expansions are
\begin{align}
E
&=e\left[1-2(\Theta\zeta)
+\frac{1}{2}\Theta^2\left(\mathcal{A}-2\zeta^2-\lambda^2\right)\right],
\label{eq:app-mp-E-bosonic-expansion}
\end{align}
and
\begin{equation}
\begin{aligned}
\mathcal R
={}&-\mathcal{A}+2(\Theta\check\gamma_*\widehat\Sigma)-2\mathcal{A}(\Theta\zeta)\\
&+\frac{1}{2}\Theta^2\left[
\frac{1}{2}\widehat R_2
+\mathcal{A}\left(\mathcal{A}-2\zeta^2-\lambda^2\right)
-4(\zeta\check\gamma_*\widehat\Sigma)
\right],
\label{eq:app-mp-R-bosonic-expansion}
\end{aligned}
\end{equation}
where the fermions $\zeta_\alpha$ and $\lambda^\mu_\alpha$ are such that $\psi^\mu_\alpha=(\zeta\check\gamma^\mu)_\alpha+\lambda^\mu_\alpha$, while
\begin{equation}
\widehat\Sigma_\alpha =\epsilon^{\mu\nu}\left( \partial_\mu\psi_{\nu\alpha} +\frac{1}{2}\widehat\omega_\mu(\check\gamma_*\psi_\nu)_\alpha \right)
\end{equation}
is the gravitino field strength, and $\widehat R_2=2\epsilon^{\mu\nu}\partial_\mu\widehat\omega_\nu.$
The hat denotes the supercovariant spin connection, including the gravitino torsion terms; in particular 
\begin{equation}
\widehat R_2=R_2+\mathcal{O}(\psi^2).
\end{equation} 
Besides the gravity sector, we also have off-shell matter multiplets $(q^I, \chi^I, F^I)$, where $q^I$ is a real scalar, $\chi^I$ a Majorana fermion and $F^I$ a real auxiliary field. In superspace, they are given by
\begin{equation}
Q^I = q^I + \frac{1}{2} \Theta \check\gamma_* \chi^I + \frac{1}{2} \Theta^2 F^I.
\end{equation}

The central object is then the superspace action
\begin{equation}
S_{SUGRA} = \int {\rm d}^2x  {\rm d}^2 \Theta \, E\, \left(J(Q)\mathcal{R}-\mathcal{K}_{IJ}(Q) \mathcal{D}^\alpha Q^I \mathcal{D}_\alpha Q^J - W(Q)\right),
\end{equation}
where $J$, $\mathcal{K}_{IJ}$ and $W$ are real functions of the matter superfields. In particular, the matrix $\mathcal{K}_{IJ}$ is symmetric and non-degenerate.  On a scalar superfield, the curved superspace derivative $\mathcal{D}^\alpha$ obeys
\begin{equation}
\{\mathcal{D}_\alpha,\mathcal{D}_\beta\} Q^I = -2 (\check\gamma^a)_{\alpha\beta}\mathcal{D}_a Q^I,
\end{equation}
and in the flat superspace limit it reduces to $\mathcal{D}_\alpha \longrightarrow D_\alpha
=\frac{\partial}{\partial \Theta^\alpha}-(\check\gamma^a\Theta)_\alpha \partial_a$, such that $\{D_\alpha,D_\beta\}=-2(\check\gamma^a)_{\alpha\beta}\partial_a$.
Each of the three terms in the superspace action is separately supersymmetric and real.

The bosonic part of the component expansion is
\begin{equation}
\begin{aligned}
S_{SUGRA} = \int d^2x \sqrt{-g_2} &\bigg( \frac{1}{2}  J(q) R_2 + 2 \mathcal{K}_{IJ}(q)\partial^\mu q^I \partial_\mu q^J \\
&- \left(2 \mathcal{K}_{IJ}(q)F^I F^J + \mathcal{A}J_I F^I + \mathcal{A}W(q)+W_I F^I\right)\bigg),
\end{aligned}
\end{equation}
where $J_I = \frac{\partial}{\partial{q^I}}J$ and $W_I = \frac{\partial}{\partial{q^I}}W$. As usual, the scalar potential is generated by integrating out the auxiliary fields $\mathcal{A}$, $F^I$. Their equations of motion give
\begin{align}
\label{eq:appauxAEOM}
\mathcal{A} &= \frac{4W - J_I \mathcal{K}^{IJ}W_J}{J_I \mathcal{K}^{IL}J_L},\\
F^I&=-\frac{1}{4} \mathcal{K}^{IL}\left(W_L + J_L\mathcal{A}\right),
\label{eq:appauxFEOM}
\end{align}
where we assumed $J_I \mathcal{K}^{IL}J_L\neq 0$ on the configuration space; this will be confirmed in our explicit model.
Substituting these expressions back into the action we obtain
\begin{equation}
\begin{aligned}
S_{SUGRA} = \int d^2x \sqrt{-g_2} \left( \frac{1}{2}  J(q) R_2 + 2 \mathcal{K}_{IJ}(q)\partial^\mu q^I \partial_\mu q^J - V+ ferm.\right),
\end{aligned}
\end{equation}
where the scalar potential is
\begin{equation}
\begin{aligned}
V&=-\frac{1}{8} \mathcal{K}^{IJ}W_I W_J + \frac{1}{8} J_I \mathcal{K}^{IL}J_L \mathcal{A}^2 \\
&= -\frac{1}{8} \mathcal{K}^{IJ}W_I W_J + \frac{1}{8}   \frac{(4W-J_I \mathcal{K}^{IL}W_L)^2}{J_I \mathcal{K}^{IL}J_L}.
\end{aligned}
\end{equation}
For completeness, we also give the supersymmetry transformations of the component fields. 
The supergravity multiplet transforms as
\begin{align}
\delta_\epsilon e_\mu{}^a
&=-2(\epsilon\check\gamma^a\psi_\mu),
\label{eq:app-mp-delta-e}\\
\delta_\epsilon\psi_\mu{}^\alpha
&=-\left[
(\widehat D_\mu\epsilon)^\alpha
-\frac{1}{2}\mathcal{A}(\epsilon\check\gamma_\mu)^\alpha
\right],
\label{eq:app-mp-delta-psi}\\
\delta_\epsilon \mathcal{A}
&=2(\epsilon\check\gamma_*\widehat\Sigma)
+\mathcal{A} e_\mu{}^a(\epsilon\check\gamma_a\psi^\mu).
\label{eq:app-mp-delta-A}
\end{align}
The matter multiplets transform as
\begin{align}
\delta_\epsilon q^I
&=-\frac{1}{2}(\epsilon\check\gamma_*\chi^I),
\label{eq:app-mp-delta-q}\\
\delta_\epsilon\chi^I_\alpha
&=-2(\check\gamma_*\epsilon)_\alpha F^I
+(\check\gamma_*\check\gamma^a\epsilon)_\alpha
(\psi_a\check\gamma_*\chi^I)
-2(\check\gamma_*\check\gamma^\mu\epsilon)_\alpha
\partial_\mu q^I,
\label{eq:app-mp-delta-chi}\\
\delta_\epsilon F^I
&=-2(\epsilon\zeta)F^I
-\frac{1}{2}
(\epsilon\check\gamma^\mu\check\gamma_*
\widehat D_\mu\chi^I)
-(\epsilon\lambda^\mu)
\left[
(\psi_\mu\check\gamma_*\chi^I)
-2\partial_\mu q^I
\right].
\label{eq:app-mp-delta-F}
\end{align}

On a background in which fermions are set to zero and scalar fields are constant, vanishing of the supersymmetry variations of the fermions reduces to 
\begin{align}
F^I&=0,\\
D_\mu\epsilon-\frac{1}{2}\mathcal{A}\check\gamma_\mu\epsilon&=0,
 \label{eq:appdpsi=0}
\end{align}
where $D_\mu \epsilon = \partial_\mu \epsilon + \frac{1}{2} \omega_\mu\check\gamma_* \epsilon$. Acting again with $D_{\nu}$ on \eqref{eq:appdpsi=0}, and using $[D_\mu, D_\nu] \epsilon= \frac{1}{4} R_2 \check\gamma_{\mu\nu}\epsilon$, we get the integrability condition
\begin{equation}
\label{eq:integrabilitySUGRA2D}
R_2 = -2 \mathcal{A}^2.
\end{equation}
This allows us to distinguish between Minkowski or AdS$_2$ vacua. 

Supersymmetric Minkowski vacua require both $F^I=0$ and $\mathcal{A}=0$. Using equations \eqref{eq:appauxAEOM} and \eqref{eq:appauxFEOM}, this is equivalent to
\begin{equation}
\label{eq:appSUSYMinkSUGRA}
\text{SUSY Minkowski} \qquad \qquad W=0, \qquad W_I=0. 
\end{equation}
Instead, supersymmetric AdS$_2$ vacua need $\mathcal{A}\neq 0$, due to \eqref{eq:integrabilitySUGRA2D}, and also $F^I=0$. Using equations \eqref{eq:appauxAEOM} and \eqref{eq:appauxFEOM}, this is equivalent to
\begin{equation}
\text{SUSY AdS$_2$} \qquad \qquad  W=0, \qquad W_I=-J_I\mathcal{A}. 
\end{equation}
Notice that the on-shell potential $V$ vanishes at both supersymmetric Minkowski and supersymmetric AdS$_2$ solutions. Indeed, in the presence of the  non-minimal coupling $J(q)R_2$, curvature is supported by the non-vanishing gravity auxiliary field rather than by a non-zero value of the scalar potential $V$.

{\small
\bibliographystyle{utphys}
\bibliography{bib}
}
\end{document}